\documentclass[preprint2]{aastex}
\usepackage{bm}
\usepackage{epstopdf}
\usepackage{adjustbox}
\usepackage{rotating}
\usepackage{pdflscape}

\def\arcsa#1#2{$#1^{\prime\prime}_{^\textrm{.}}#2$}
\def\arcs#1{$#1^{\prime\prime}$}

\shorttitle{The Quadrupolar PPN IRAS 17150-3224 }
\shortauthors{Huang, Lee, \& Sahai}

\begin{document}

\title{Evolution from Spherical AGB Wind to Multipolar Outflow in Pre-Planetary Nebula IRAS 17150-3224}

\author{Po-Sheng Huang\altaffilmark{1,2}, Chin-Fei Lee\altaffilmark{2,1}, \& Raghvendra Sahai\altaffilmark{3}}


\altaffiltext{1}{Institute of Astrophysics, National Taiwan University,
Taipei 106, Taiwan ; posheng@asiaa.sinica.edu.tw}

\altaffiltext{2}{Institute of Astronomy and Astrophysics, Academia Sinica, P.O. Box 23-141, Taipei 106, Taiwan}
\altaffiltext{3}{Jet Propulsion Laboratory, MS 183-900, California Institute
of Technology, Pasadena, CA 91109, USA}

\begin{abstract}

We have mapped  the pre-planetary nebula IRAS 17150-3224 in 350 GHz
continuum and CO $J=3$--2 line at an angular resolution of $\sim$
\arcsa{0}{09} using the Atacama Large Millimeter/submillimeter Array.  A
continuum source is detected at the center of the nebula, elongated along
the equatorial plane, likely tracing a dusty torus around the central
source.  Continuum emission is also detected on both sides of the central
continuum source in the equatorial plane, probably resulting from
interactions of collimated fast winds with envelope material in the equator. 
CO emission is detected along the optical lobe.  Although the optical lobe
appears as bipolar, CO map shows that it is actually a quadrupolar outflow
consisting of two overlapping bipolar outflows.  Two additional younger
bipolar outflows are also detected in CO, one at lower latitude and the
other along the equatorial plane.  In the CO position-velocity maps,
blueshifted absorption stripes are detected in the outflow emissions, due to
absorption by a series of shells produced by a series of asymptotic giant
branch (AGB) wind ejections.  By modeling the morphology and kinematics of
the AGB wind and outflows, we find that the AGB wind could end $\sim$ 1300
yr ago, the quadrupolar outflow was ejected $\sim$ 350 yr ago, and the two
additional bipolar outflows were ejected $\sim$ 280 and 200 ago,
respectively.  The outflows could be produced either by bullets coming 
from an explosion, or by a precessing collimated fast wind 
with a time-dependent ejection velocity.

\end{abstract}



\clearpage

\section{Introduction}

Pre-planetary nebula (PPN) phase is a short ($\sim$ 1000 yr) evolutionary
phase between the asymptotic giant branch (AGB) phase and the planetary
nebula (PN) phase for low- to intermediate-mass stars.  Most PPNs have
developed aspherical morphologies in a few hundred years, possessing
toroidal envelope and collimated outflows \citep{ST98,B01,S07}.  As a result,
PPNs are considered to be the key to answering how spherical AGB winds
transform to aspherical PNs \citep[see, e.g.,][]{BF02}.  Recent observations
have shown that bipolar PPNs, e.g., the red rectangle and IRAS 08544-4431
\citep{B16,B18}, could be shaped by a disk wind launched from a rotating
disk.  On the other hand, multipolar PPNs, e.g., CRL 618, could be shaped
either by bullets \citep{B13,L13a,H16} or by a precessing collimated fast
wind and jet with time-dependent ejection velocity \citep{R14,V14}.  
The bullets could come from explosive events,
producing multipolar outflows.  The outflows should have the same age, if
the bullets are produced simultaneously.  On the other hand, the precessing
collimated fast wind and jet could come from a precessing disk, producing
multipolar outflows at different angles at different ages.

Here we present the Atacama Large Millimeter/submillimeter Array (ALMA)
observations of the PPN Cotton Candy Nebula (IRAS 17150-3224; hereafter
I17150) at high angular resolution in 350 GHz continuum and CO $J=3$--2
line.  In optical, I17150 appears as a bipolar PPN based on its optical
image obtained by the $Hubble$ $Space$ $Telescope$ ($HST$)\citep{K98}.  It
has a pair of highly extended optical lobes, crossed by many arc structures. 
It was also observed in CO $J=1$--0 line \citep{ZD86,H93} and $J=2$--1 line
\citep{H93}.  The central star is an OH$/$IR star with a luminosity
$L=27200$ L$_{\sun}$ and an effective temperature $T_{\rm{eff}}=5200$ K
\citep{M02}.  \citet{H93} derived a systemic velocity $V_{\rm{sys}}=14$ km
s$^{-1}$ from the OH maser and the CO $J=2$--1 observations.  The distance
to I17150 is not well constrained.  \citet{D05} assumed a distance of 2.42
kpc.  \citet{M02} assumed a distance of 3.6 kpc by fitting the galactic
H{\scriptsize I} velocity curve with a systemic velocity $V_{\rm{sys}}$ of
15 km s$^{-1}$.  In this paper, we assume $d=3$ kpc as an average and
$V_{\rm{sys}} = 14$ km s$^{-1}$.
We describe our ALMA observations of I17150 in sections 2 and 3; 
Our model is described in section 4; Discussions are in section 5; 
Summary of this work is in section 6.

\section{Observations}

Observations of I17150 were carried out with ALMA in band 7 with 47 antennas
in configuration of C40-6 on 2017 August 27.  The shortest baseline was
$\sim$ 38 m, and the longest baseline was $\sim$ 3.55 km.  The target was
observed for $\sim$ 50 min.  The quasars J1924-2914, J1733-1304, and
J1717-3342 were observed for the bandpass, flux, and phase calibration,
respectively.  The channel width is $\sim$ 244 kHz, giving a velocity
resolution of $\sim$ 0.212 km s$^{-1}$ per channel.  The calibration and
imaging were done by the ALMA pipeline in version r39732.

For the 350 GHz continuum map, a weighting with a robustness parameter of
0.5 was used for the uv visibilities.  The synthesized beam has
a size of $\sim$ \arcsa{0}{09} $\times$ \arcsa{0}{06} with a position
angle (P.A.) of $\sim$ $-$81$^{\circ}$.  The noise level is $\sim$
0.1 mJy beam$^{-1}$.

For the CO $J=3$--2 maps, a weighting with a robustness parameter of 0.5 was
adopted for the uv visibilities.  The synthesized beam has a size of $\sim$
\arcsa{0}{09} $\times$ \arcsa{0}{06} with a P.A. of $\sim -$76$^{\circ}$.  The
noise level in the representative spectral window of CO is $\sim$ 1.7 mJy
beam$^{-1}$ for a channel width of 0.212 km s$^{-1}$.

\section{Results}

\subsection{350 GHz Continuum}

\begin{figure*}
\centering{
\hspace{-1cm}
\includegraphics[angle=-90,scale=0.7]{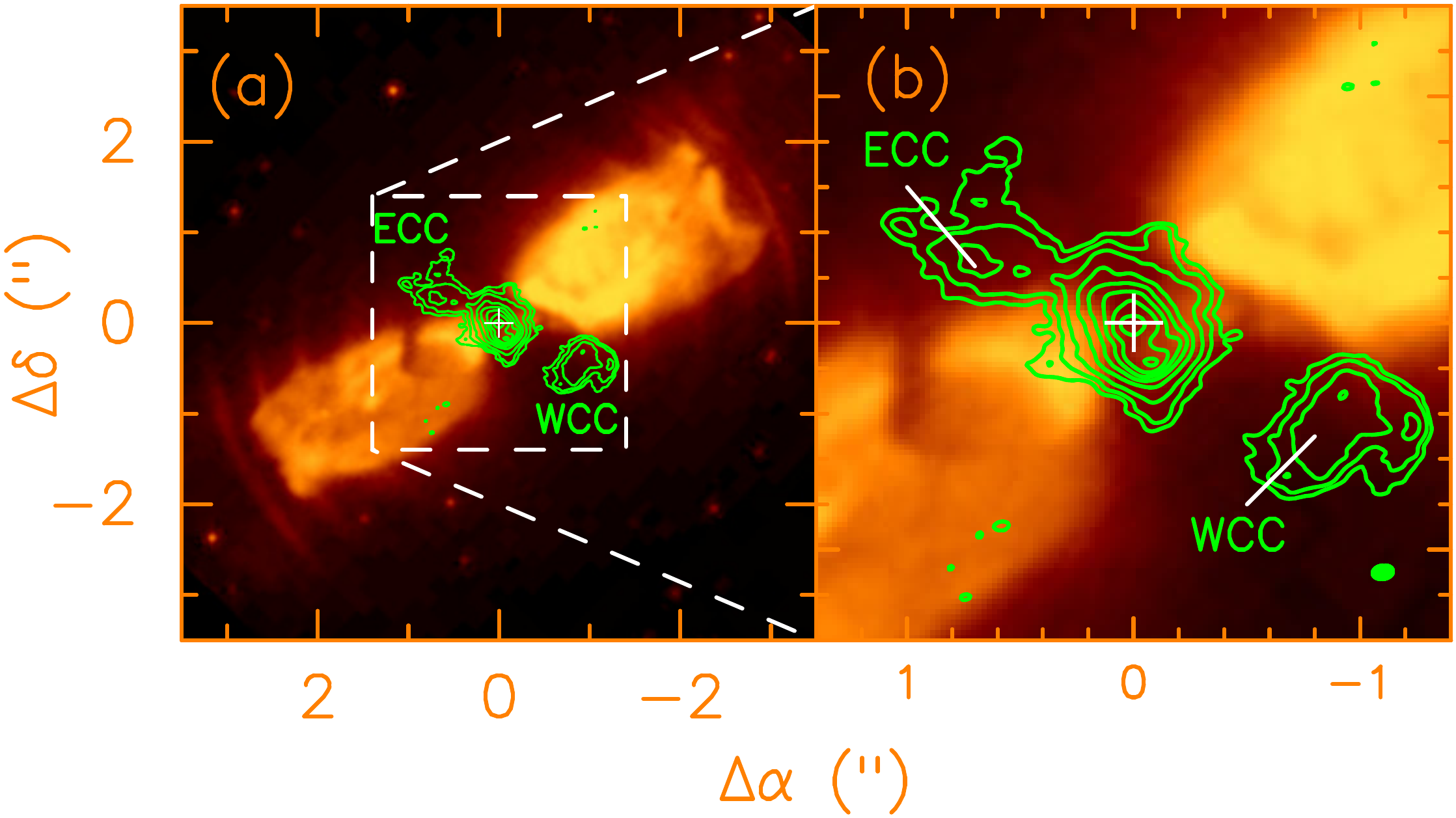}
\caption{
(a) The 350 GHz continuum map (green contours) of I17150 observed by the ALMA,
superimposed on the {\it HST} image obtained with filter F606W \citep{K98}.  
In continuum, three components, with one at the center, one in the east,
and one in the west, are detected in the
waist of the optical bipolar lobe.  (b) A zoom-in to the inner region.
The cross marks the primary peak of the central component.
The beam in the bottom-right corner has a size of
$\sim$ \arcsa{0}{9} $\times$ \arcsa{0}{06} with a major axis at
P.A. $\sim$ $-$81\degr .  The contour levels are 3$\sigma$, 
6$\sigma$, and 12$\sigma$ for the first three contours and have
a step of 12$\sigma$ for the other contours, where $\sigma \simeq 0.1$ mJy beam$^{-1}$.
\label{F1}}
}
\end{figure*}

Figure \ref{F1} shows the continuum map of I17150 at 350 GHz, superimposed
on the optical image taken by the {\it HST} with filter F606W \citep{K98}. 
Three continuum components are detected in continuum, with one at the
center, one in the east (ECC), and one in the west (WCC) in the equatorial
plane.  They have a total extent of $\sim$ \arcsa{2}{6} ($\sim$ 7800 au)
along the equator.  The central component is elongated along the equatorial
plane with a primary peak at the center of the nebula (Figure \ref{F1}b) and
thus likely traces a torus around the central source of I17150, as included
in the circumstellar envelope model proposed in \citet{M02}.  The ICRS
coordinate of the primary peak is R.A. 
$=17^{\rm{h}}18^{\rm{m}}19^{\rm{s}}_{^.}875$, Dec.  $=-32$\degr \,
$27^{\prime}$ \arcsa{21}{8}.  A secondary peak is seen at a distance of
$\sim$ 550 au (\arcsa{0}{18}) to the southwest of the primary peak and seems
to be spatially aligned with the linear optical emission there in the dust
lane.  The eastern and western components are located at a large distance of
$\sim$ \arcs{1} ($\sim$ 3000 au) from the central component, and thus could
trace materials produced and dragged away by winds or jets from the central
source, as discussed later.

The total flux density of the continuum in the three components is $\sim$
0.3 Jy, amounted to $\simeq$ 40\% of the value ($\sim$ 0.8 Jy) previously
reported at a similar frequency in \citet{H93} with James Clerk Maxwell 
Telescope (JCMT) at a resolution of $\sim$ \arcs{19}.  This suggests that 
although the three components here are compact, they have a flux density 
comparable to that of the extended component (AGB wind).  
Assuming that the continuum emission is all thermal dust emission and 
optically thin, the dust mass producing the continuum emission can be 
roughly estimated from the following equation:
\begin{equation} M_{d} \simeq \frac{D^2 F_{\nu} }{B_{\nu} \left( T_{d}
\right) {\kappa}_{\nu} }, \end{equation} where $D$ is the distance to
I17150, $F_{\nu}$ is the flux density, $B_{\nu} \left( T_{d}\right)$ 
is the Plank blackbody function 
at the dust temperature $T_{d} \sim 150$ K \citep{M02,dV15}, and
${\kappa}_{\nu}$ is the dust grain opacity.  Assuming that the dust grains
are similar to to those around evolved stars \citep{D18}, we have
${\kappa}_{\nu}=0.3$ cm$^2$ g$^{-1}$ at 350 GHz, and thus $M_{d} \sim 7
\times 10^{-2}$ M$_{\sun}$.

The gas-to-dust mass ratio of the circumstellar envelope of I17150 is 
unknown.  Previously, \citet{M02} adopted a ratio of 280 as in the shell of 
the OH/IR star OH 26.5+0.6 \citep{J96}.  Using this ratio, we derive a shell 
mass of $\sim$ 20 M$_{\sun}$, which seems too big for an OH/IR star, like 
our source.  Thus, we adopt a ratio of 77, as found in the 
circumstellar envelope around O-rich stars \citep{D18} to calculate the mass 
and find it to be $\sim$ 5.4 M$_{\sun}$, which is more reasonable
for an OH/IR star.

\subsection{CO Intensity and Velocity Maps}

\begin{figure*}
\centering{
\includegraphics[angle=-90,scale=1]{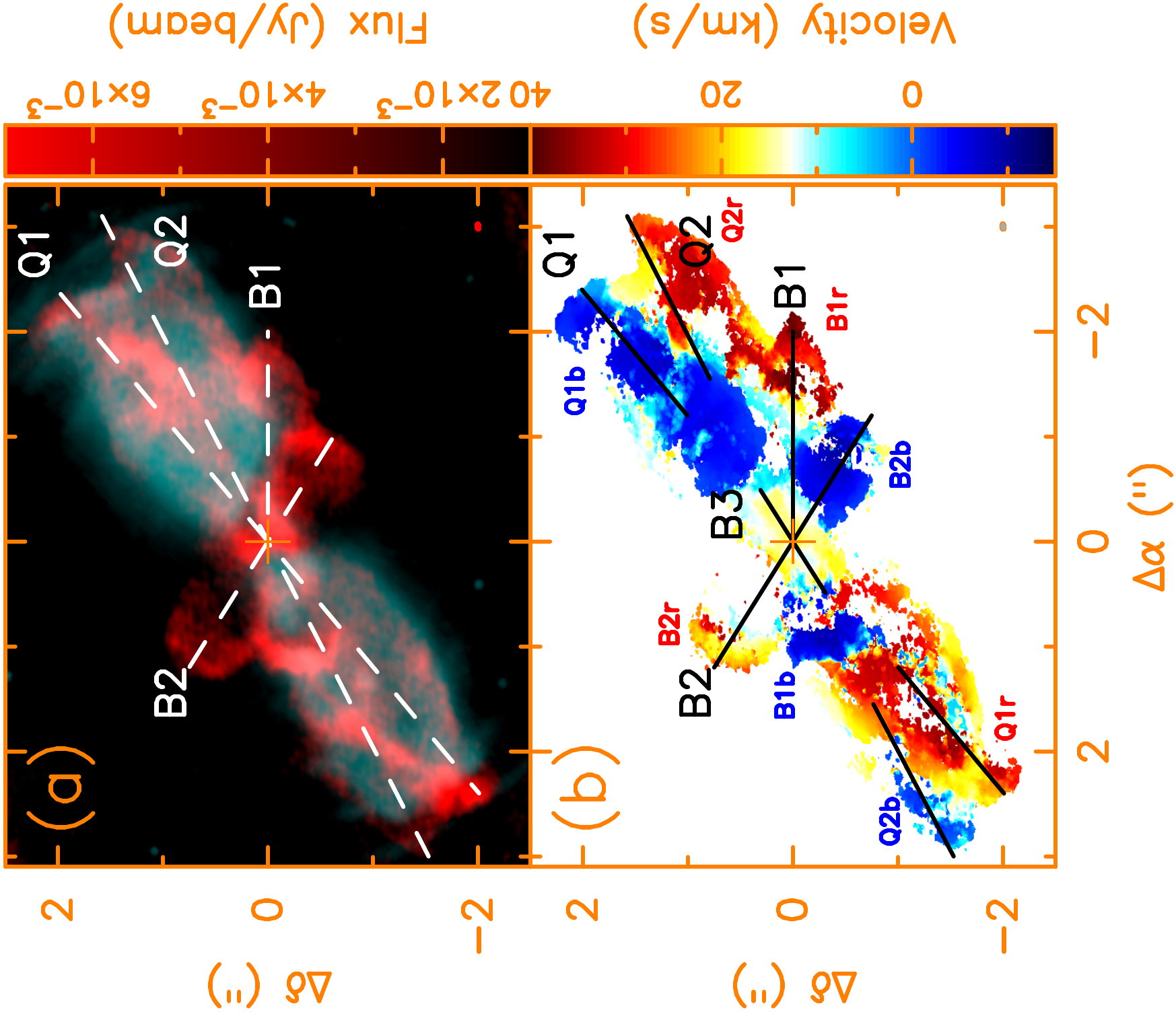}
\caption{
(a) Total intensity map (red image) plotted on top of the {\it HST} image 
(turquoise image) and (b) intensity-weighted velocity map of I17150 in 
CO $J=3$--2 line.  The dashed lines and solid lines indicate the outflow axes. 
Blueshifted and redshifted outflow lobes are labeled with symbols b and r, 
respectively.  The low-velocity emission at the center (yellow color) 
is from outflow B3 as indicated by the outflow axis.  
The beam size is $\sim$ \arcsa{0}{09} $\times$ \arcsa{0}{06} 
with a major axis at P.A. $\sim$ $-$81\degr .
\label{F2}}
}
\end{figure*}

Figure \ref{F2}a shows the total intensity map (red image) of the CO
$J=3$--2 emission integrated from $-$15 to 45 km s$^{-1}$, superimposed on
the {\it HST} image (turquoise image).  It shows a quadrupolar outflow
consisting of two overlapping bipolar outflows, Q1 and Q2, along
the optical lobes.  It also shows two additional bipolar outflows, B1 at a
lower latitude and B2 along the equator, that have not been detected before.

Figure \ref{F2}b shows the intensity-weighted (radial) velocity map toward
the nebula, in order to identify different velocity components of the
outflows.  The solid lines indicate the axes of the quadrupolar outflow Q1
and Q2, and the bipolar outflows B1 and B2.  As can be seen, outflow Q1 has
a blueshifted lobe (Q1b) in the northwest and a redshifted lobe (Q1r) in the
southeast, whereas outflow Q2 has a blueshifted lobe (Q2b) in the southeast,
and a redshifted lobe (Q2r) in the northwest.  Thus, Q1 and Q2 have opposite
inclination angles to the plane of the sky.  The length and width of the
lobes in Q1 and Q2 are $\sim$ \arcsa{3}{2} (9600 au) and \arcs{1} (3000 au),
respectively.  At a lower latitude, outflow B1 has a redshifted lobe (B1r)
in the west.  In the east, the blueshifted lobe (B1b) roughly along the Q2b
axis is considered here to be the blueshifted counterpart, although it appears
deflected from the B1 axis by $\sim$ 17\degr\ counterclockwise.  The length
of lobe B1b is $\sim$ \arcs{1} (3000 au), about half of lobe B1r, which is
$\sim$ \arcs{2} (6000 au).  Outflow B2 has a blueshifted lobe (B2b) in the
southwest, and a redshifted lobe (Q2r) in the northeast along the equator,
aligned with the major axis of the continuum emissions.  The length and
width of lobes B2b and B2r are $\sim$ \arcsa{1}{4} (4200 au) and
\arcsa{0}{8} (2400 au), respectively.  Most of the outflows are reasonably
symmetric, except for outflow B1.

\subsection{CO Line Profile}

\begin{figure*}
\centering{
\hspace{-2cm}
\includegraphics[angle=-90,scale=0.6]{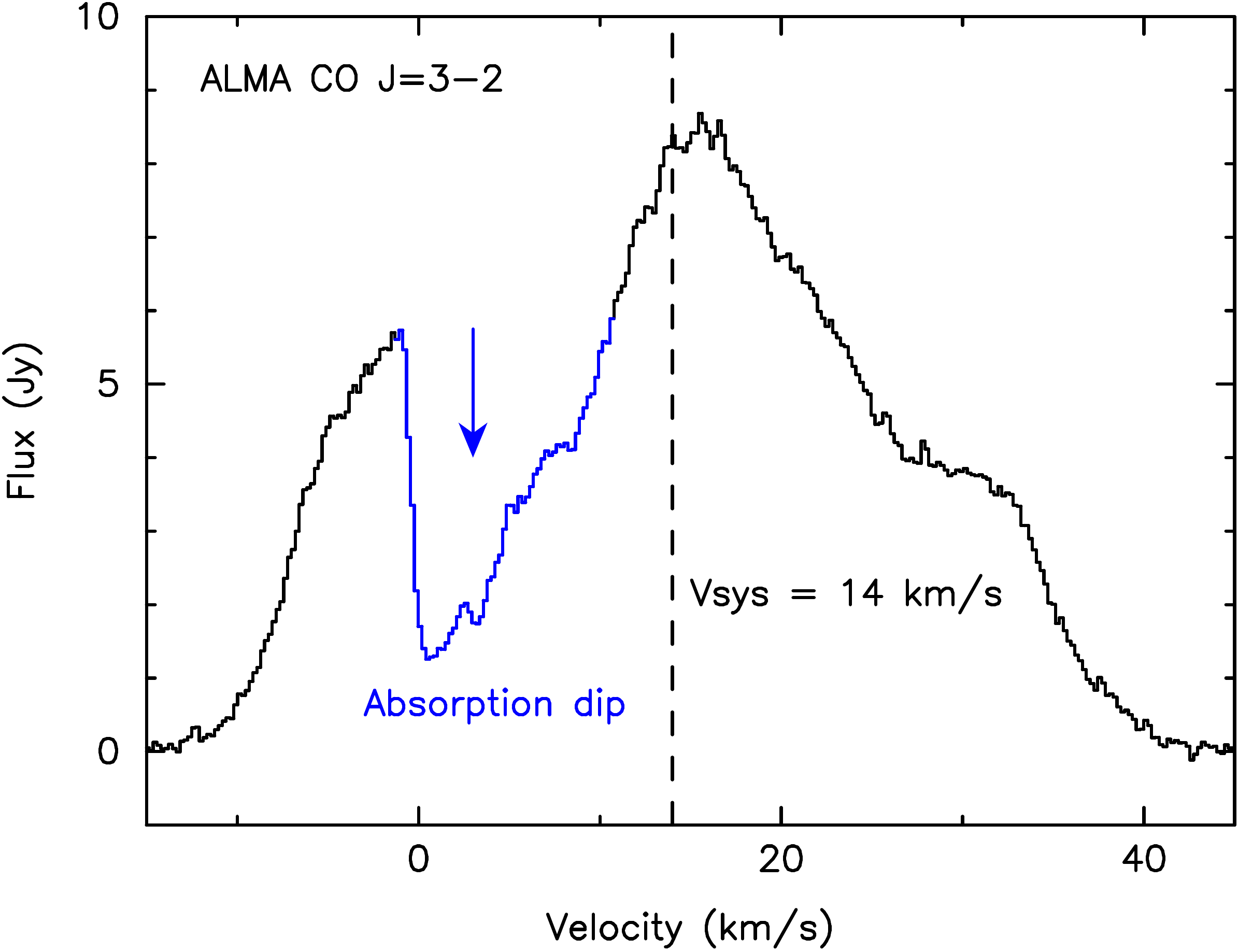}
\caption{
The CO $J=3$--2 line profile of I17150, averaged over an
elliptical region with a major axis of \arcs{7}, a minor axis of \arcs{4},  
and P.A.  = $-$56\degr, covering all the outflows.
Blue curve marks the absorption dip
at the velocity from $-$1 to 12 km s$^{-1}$. 
\label{F3}}
}
\end{figure*}

Figure \ref{F3} shows the CO $J=3$--2 line profile averaged over an
elliptical region with a major axis of \arcs{7}, a minor axis of \arcs{4},
and P.A.  = $-$56\degr, covering all the outflows.  The emission is detected
from $-$15 to 45 km s$^{-1}$.  The line profile has a peak flux of $\sim$
8.6 Jy at $\sim$ 16 km s$^{-1}$, which is $\sim$ 2 km s$^{-1}$ redshifted
from the systemic velocity.  It appears asymmetric about the systemic
velocity, mainly due to an absorption dip from $\sim$ $-$1 to 12 km s$^{-1}$
produced by the front part of a cooler AGB wind, as discussed in Section
\ref{PV}.

\subsection{CO Channel Maps} \label{CH}

\begin{figure*}
\centering{
\hspace{-1cm}
\includegraphics[angle=-90,scale=0.7]{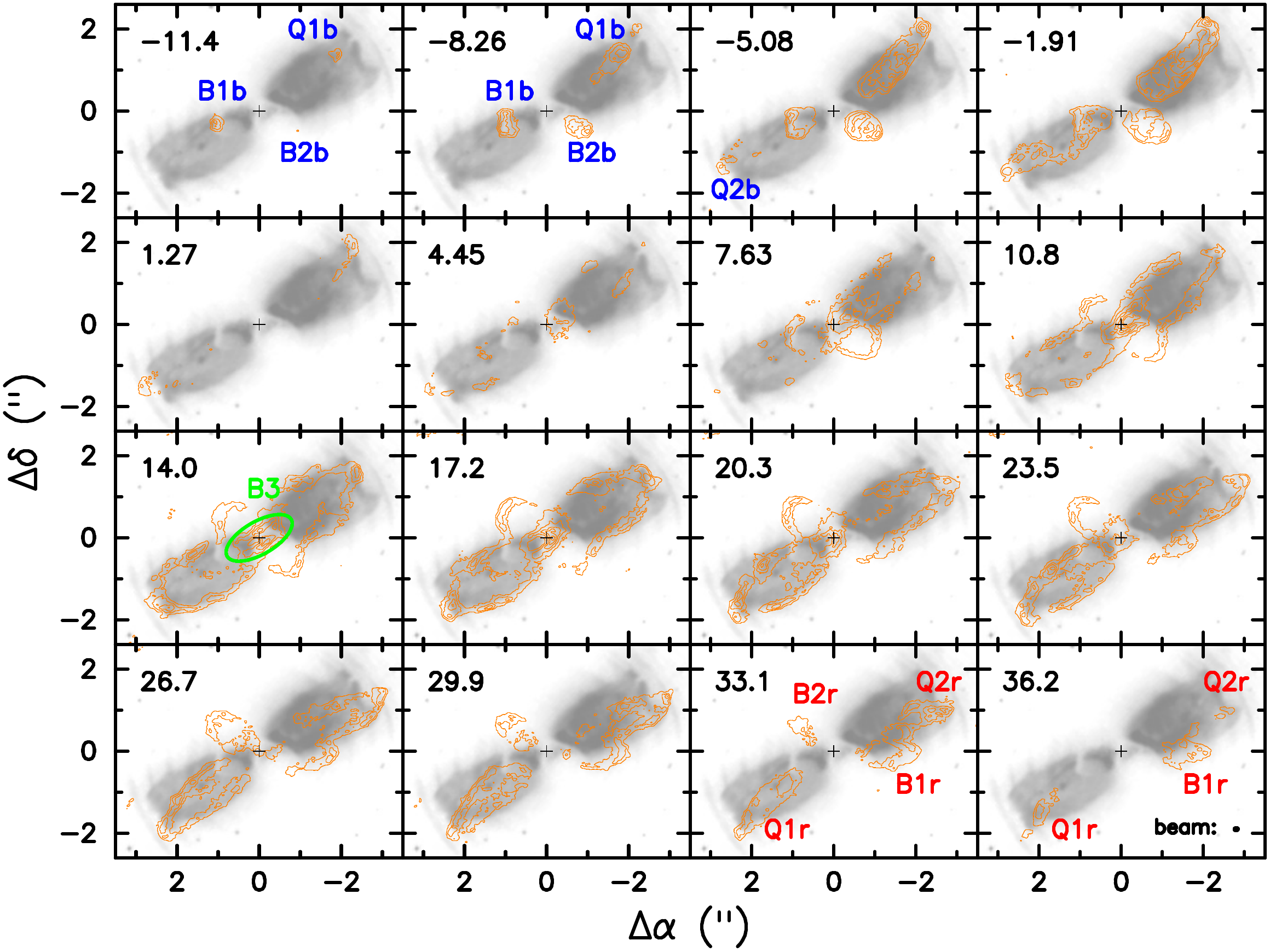}
\caption{
The CO $J=3$--2 channel maps of I17150. The velocity in each channel
is indicated at the upper-left corner. The systemic velocity is $\sim$ 14 
km s$^{-1}$.  The beam size is $\sim$ \arcsa{0}{09} $\times$ \arcsa{0}{06} 
with a major axis at P.A. $\sim$ $-$81\degr .  The contours start from 3$\sigma$ 
with a step of 3$\sigma$, where $\sigma \simeq 3$ mJy beam$^{-1}$.  
The quadrupolar outflow lobes are labeled with Q1b, Q1r, Q2b, and Q2r.  
The bipolar outflow lobes are labeled with B1b, B1r, B2b, and B2r.  
The outflow at the base of the outflows is labeled with B3. 
\label{F4}}
}
\end{figure*}

The detailed velocity distribution of the outflow lobes can be studied with
the CO $J=3$--2 channel maps shown in Figure \ref{F4}.  In order to increase
the signal-to-noise ratio to better see the velocity distribution, we binned
15 channels to one single wide channel with a velocity width of $\sim$ 3.2
km s$^{-1}$.  The outflow lobes can be better distinguished at high
velocities.  For example, four blueshifted outflow lobes Q1b, Q2b, B1b, and
B2b are clearly seen in high blueshifted-velocity channels at $-$11.4,
$-$8.26, and $-$5.08 km s$^{-1}$, and four redshifted outflow lobes Q1r,
Q2r, B1r, and B2r are clearly seen in high redshifted-velocity channels at
33.1 and 36.2 km s$^{-1}$.  At low velocity around the systemic velocity
(14.0 km s$^{-1}$), the CO emission traces the limb-brightened cavity walls
of the outflows.  Interestingly, the cavity walls of the 4 lobes (Q1b, Q1r,
Q2b, and Q2r) in the quadrupolar outflow merge side by side, appearing as a
bipolar outflow around the optical bipolar lobe.  This is probably because
the outflow lobes interact with each other, and the CO molecules in the
interacting zones have been dissociated by the interaction.

From the systemic-velocity channel (at 14.0 km s$^{-1}$) to the
high-velocity channels (to $-$11.4 km s$^{-1}$ on the blueshifted side and
to 36.2 km s$^{-1}$ on the redshifted side), the CO emission of the
quadrupolar outflow shows lobe structures that narrow down to the outflow
axis and shrink towards the outflow tips.  Since the lobe structure in each
channel map comes from a sectional view of the outflow lobes, this indicates
that the outflow lobes are hollow and ellipsoidal and has a velocity
increasing towards the tips.  At the systemic-velocity channel, the outflow
shells can be seen clearly and have a thickness of $\sim$300 au.  In the
velocity range from 1.27 km s$^{-1}$ to 7.63 km s$^{-1}$, the CO emission is
weaker than that in the other velocity range, because the emission is mostly
absorbed by a front part of the AGB wind, as implied from the absorption
dip in the line profile shown in Figure \ref{F3}.

\begin{figure*}
\centering{
\includegraphics[angle=-90,scale=0.8]{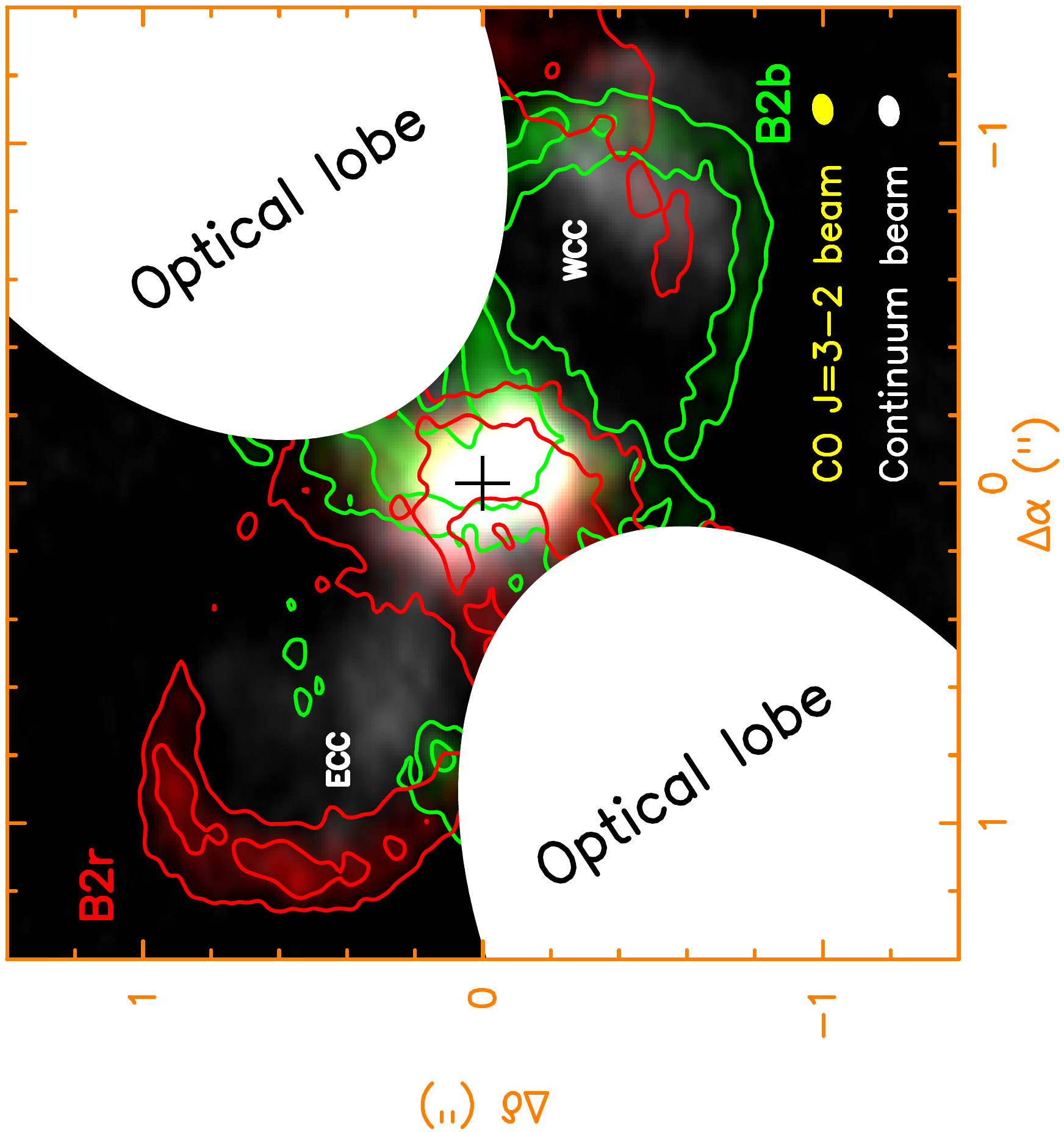}
\caption{
The morphological connection between
the continuum and outflows in the inner region of I17150. It shows
the blueshifted (at 7.63 km s$^{-1}$, green image and contours) 
and redshifted (at 20.3 km s$^{-1}$, red image and contours) CO channel maps 
of the outflows, superimposed on the continuum map (grey-scale image).   
The outflow B2 is roughly aligned with the axis connecting the eastern and 
western continuum components.  The contour levels of the CO emissions
are the same as those in Figures \ref{F4}.
\label{F5}}
}
\end{figure*}

Outflow B2 is along the equatorial plane, roughly aligning with the axis
connecting the eastern (ECC) and western continuum components (WCC), as
shown in Figure \ref{F5}.  ECC is inside the redshifted lobe B2r, extending
towards its tip.  WCC has two parts, with the inner part located right
inside the tip of the blueshifted lobe B2b, and the outer part right outside
the tip.  ECC and WCC could result from interactions between the outflow and
AGB wind or torus in the equator.

\begin{figure*}
\centering{
\includegraphics[angle=-90,scale=0.7]{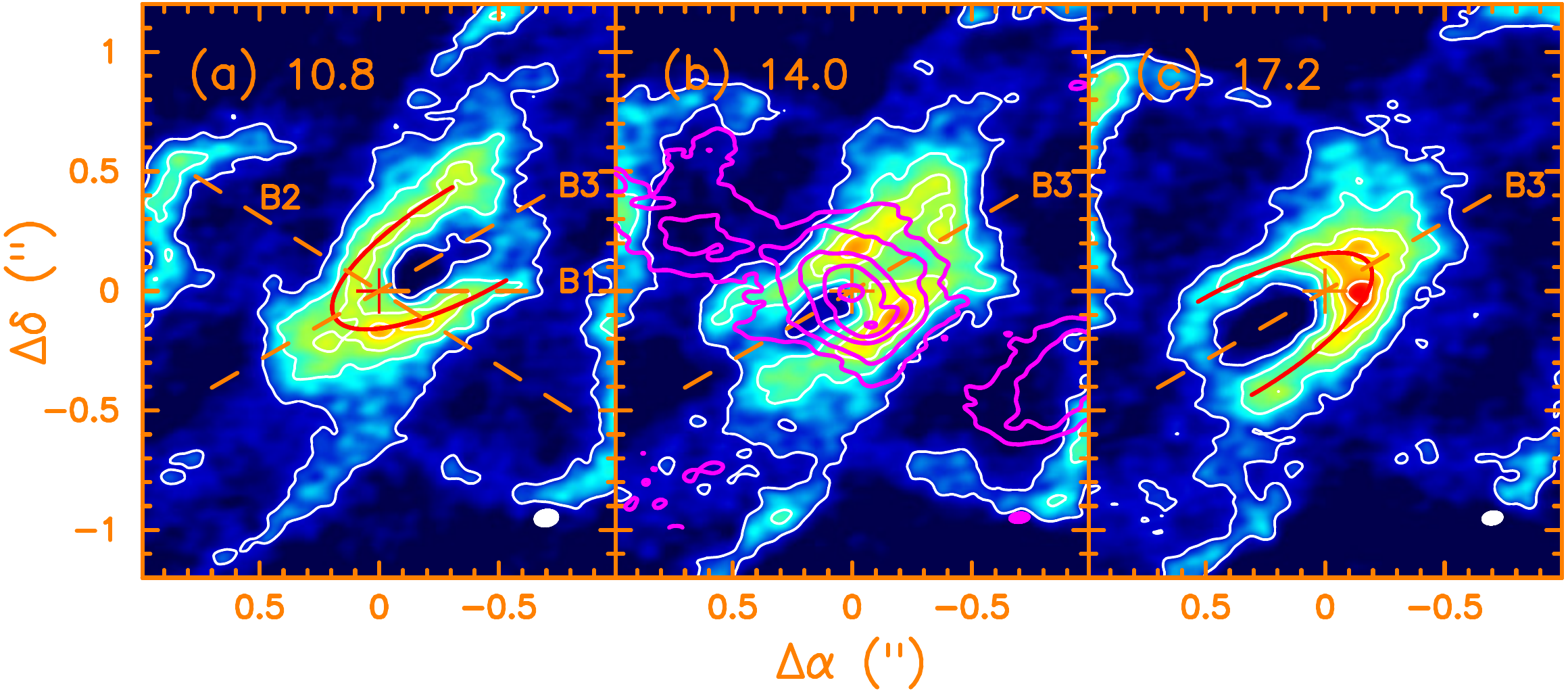}
\caption{
The low-velocity CO channel maps of outflow B3 in I17150.  
The velocity in each channel is indicated in the upper-left corner.
Panels (a) and (c) show the low-velocity
blueshifted and redshifted shells, respectively, as indicated by the red curves (which
are the fits using a half ellipse).
Panel (b) shows the continuum map (magenta contours) superimposed on the CO map of
the outflow at the systemic velocity of 14 km s$^{-1}$, in order to reveal
their morphological connection. The beam and contour levels
of the CO maps are the same as those in Figure \ref{F4}. 
\label{F6}}
}
\end{figure*}

At low velocity, as in the channels of 7.63--23 km s$^{-1}$ in Figure
\ref{F4}, when we look into the very center within \arcsa{0}{7} of the
central source, we see an additional smaller bipolar structure extending out from the
source along the symmetry axis of the optical lobes (P.A.  = $-$64\degr{}). 
Figure \ref{F6} shows the zoom-in in the central low-velocity channels of
10.8--17.2 km s$^{-1}$ in order to better study the their structures.  The
blueshifted (Figure \ref{F6}a) and redshifted (Figure \ref{F6}c) emissions
show a pair of U-shaped cavity walls, which have a total length of $\sim$ 1800 au
(\arcsa{0}{8}) and a width of $\sim$ 1800 au (\arcsa{0}{6}).  The cavity walls have a
symmetry center at
 the primary peak of the continuum emission (Figure \ref{F6}b) and are thus
produced by the central source, tracing the walls of an outflow (hereafter
outflow B3) near the central source.

\subsection{Position-Velocity Diagrams} \label{PV}

\begin{figure*}
\centering{
\includegraphics[angle=-90,scale=0.7]{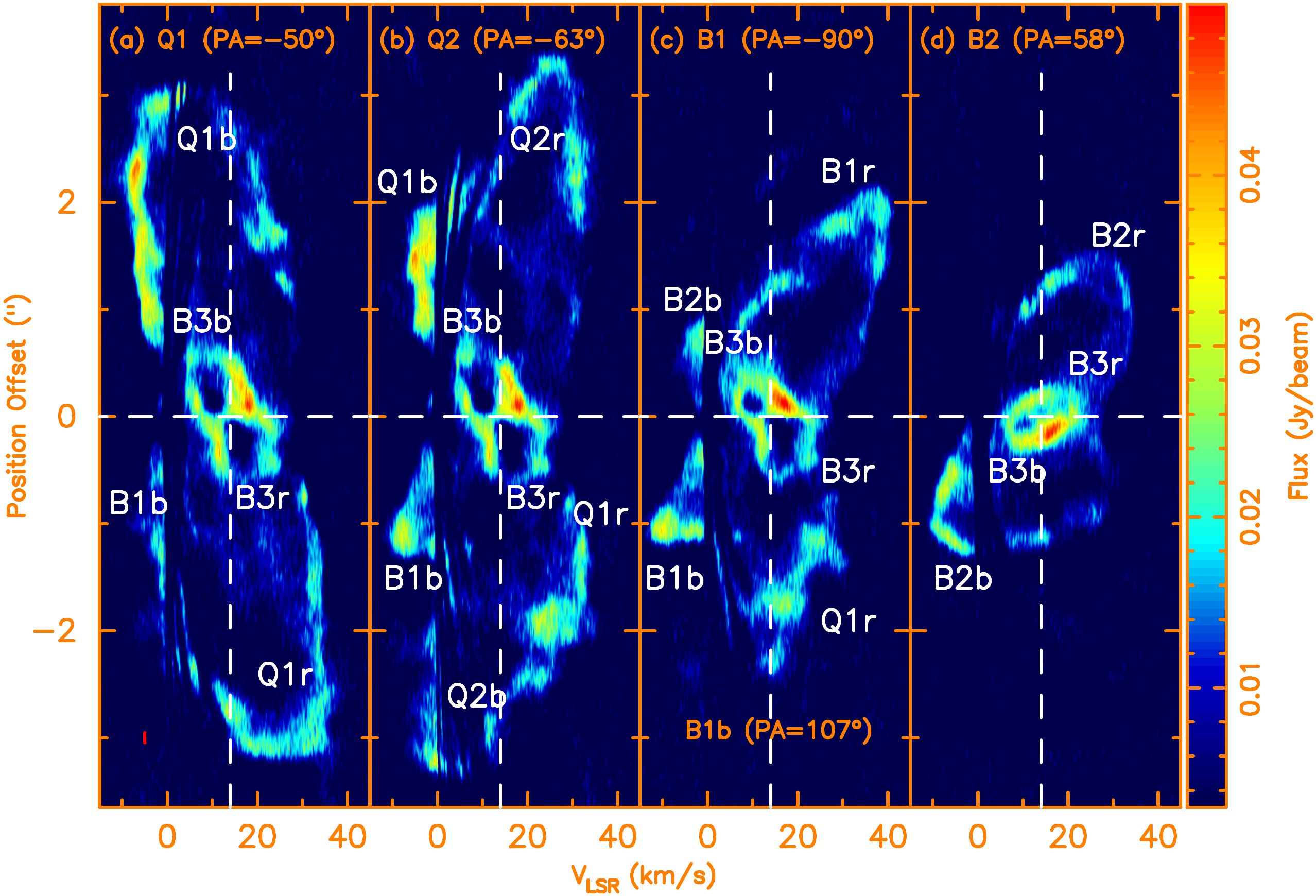}
\caption{
PV diagrams of CO $J=3$--2 emissions cut along the axes of outflows
(a) Q1, (b) Q2, (c) B1 (B1r and B1b), and (d) B2.  Vertical dashed lines
indicate the systemic velocity of 14 km s$^{-1}$ and horizontal dashed lines
indicate the central source position.  The red bar at the bottom-left corner 
indicates the resolution for the PV diagram. 
\label{F7}}
}
\end{figure*}

Position-velocity (PV) diagrams can be used to reveal the kinematics of the
outflow lobes.  Figure \ref{F7}a, \ref{F7}b, \ref{F7}c, and \ref{F7}d 
show the PV diagrams of the CO $J=3$--2 emissions cut along the axes of 
outflows Q1, Q2, B1, and B2, respectively.  
The position of the central source (\arcs{0}) and systemic velocity 
(14 km s$^{-1}$) are indicated by horizontal and vertical
dashed lines, respectively.  In Figure \ref{F7}a, lobes Q1b and Q1r are seen
with two big tilted elliptical PV structures, as expected if they are
expanding shells driven by collimated fast winds (CFWs) \citep{L01}. 
Interestingly, on the blueshifted side, a series of slightly bended
absorption stripes with a velocity ranging from $\sim$ 0 to 12 km s$^{-1}$
are detected.  They are mainly vertical but bended slightly towards the
systemic velocity.  As discussed later, they are due to the absorption by a
series of dense AGB shells in front of the outflow lobes.  Within \arcs{1} of
the central source, a smaller bipolar PV structure is seen coming from
outflow B3 lying along the axis of outflow Q1.

In Figure \ref{F7}b, two similar large elliptical PV structures are seen
coming from lobes Q2r and Q2b, but with an opposite tilt because of an
opposite inclination.  Again, the smaller bipolar PV structure comes from
outflow B3 that lies along the outflow axis of Q2.  In addition,
emissions (as marked as Q1b, Q1r, and B1b) are also seen from outflows Q1
and B1 that lie along the outflow axis of Q2.

Figure \ref{F7}c shows PV structures along the B1 (B1r) axis and the 
B1b axis (which is deflected by 17\degr\ from the B1 axis).  
The length of lobe B1r is nearly twice the length of lobe B1b.  The B1b axis 
also crosses lobe Q1r in the south, showing some emission from 
lobe Q1r in the lower part of the PV diagram.  The emissions at the center are 
from outflow B3, similar to those seen in Figure \ref{F7}a and \ref{F7}b.

In Figure \ref{F7}d, the emissions from lobes B2r and B2b form similar
elliptical PV structures and collimations to those seen from the quadrupolar
outflow lobes, and thus could also be shaped by a CFW.  In this diagram, the
cut is roughly along the equatorial plane crossing the base of outflow B3. 
The emission there forms a single elliptical lobe structure, indicating that
the base of outflow B3 is expanding equatorially.

\subsubsection{Expansion Velocity of the AGB Wind}

\begin{figure*}
\centering{
\includegraphics[angle=-90,scale=0.75]{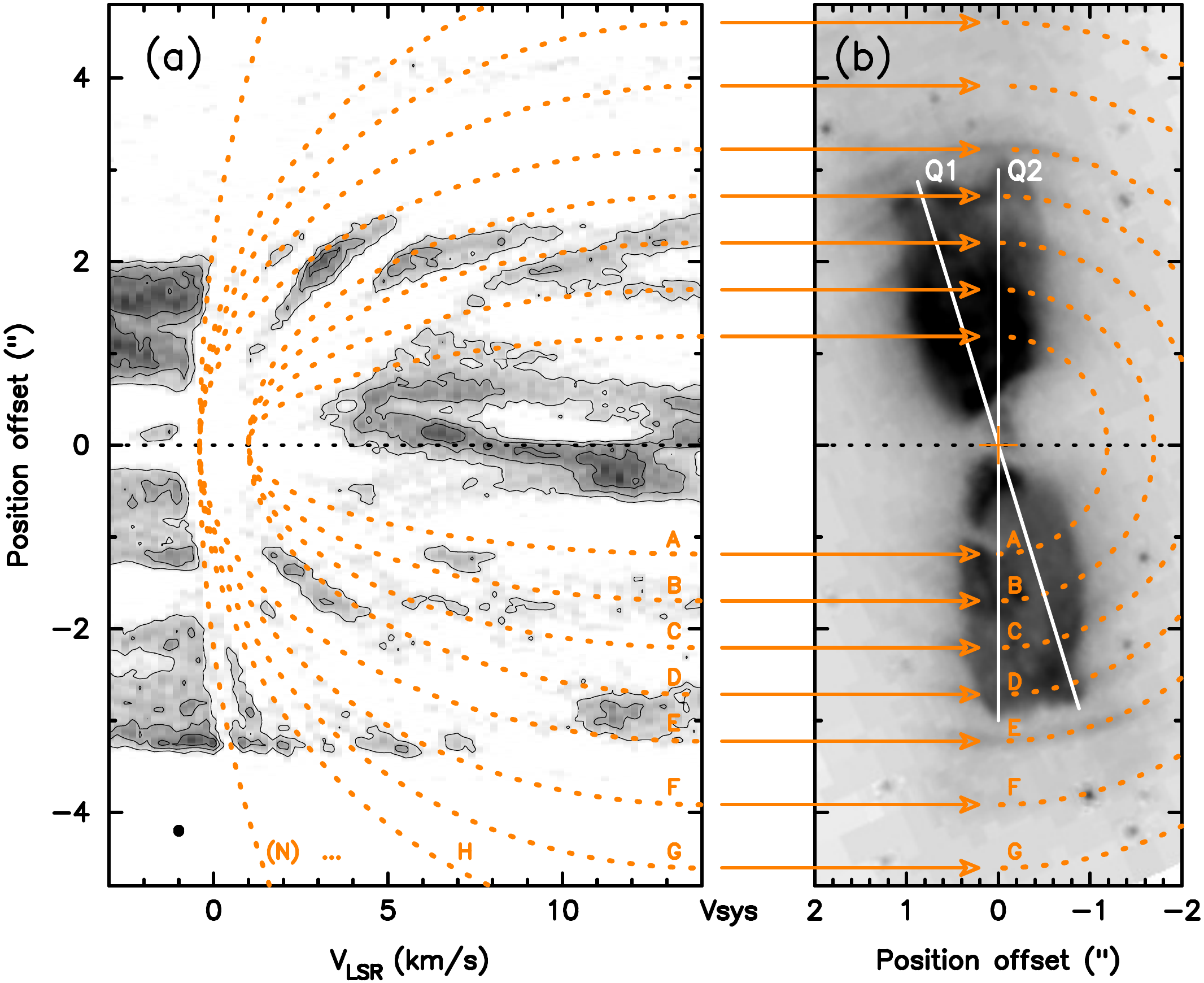}
\caption{ 
The connection of the CO PV diagram with the optical arcs.
(a) shows the PV diagram of CO cut along the axis of outflow Q2,
as extracted from the blueshifted part of Figure \ref{F7}b. 
The contours start from 3$\sigma$ with a step of 3$\sigma$, where
$\sigma \simeq 3$ mJy beam$^{-1}$. The dashed lines show the absorption curves.  
These curves can be linked to the optical 
arcs seen in the $HST$ image \citep{K98} shown in panel (b). 
\label{F8}}
}
\end{figure*}

The absorption stripes can be linked to the arc structures in the optical
image, which are due to a periodically enhanced mass loss of the
AGB wind \citep{K98}, allowing us to estimate the expansion velocity of the
AGB wind.  Previously, seven arc structures, A, B, C, D, E, F, and G, have
been identified at a distance of $\sim$ \arcsa{1}{2}, \arcsa{1}{7},
\arcsa{2}{2}, \arcsa{2}{7}, \arcsa{3}{2}, \arcsa{3}{9}, and \arcsa{4}{6},
respectively, as shown in Figure \ref{F8}b, tracing a series of dense
spherical expanding shells.  Their emission is expected to produce a series
of elliptical PV curves centered at the systemic velocity in the PV diagram. 
However, since they are cooler than the outflows, the outflow emissions are
absorbed by the front part of the AGB shells in front of the outflows,
producing the absorption stripes only on the blueshifted side, as seen in
the PV diagram.

\begin{figure*}
\centering{
\includegraphics[angle=-90,scale=0.7]{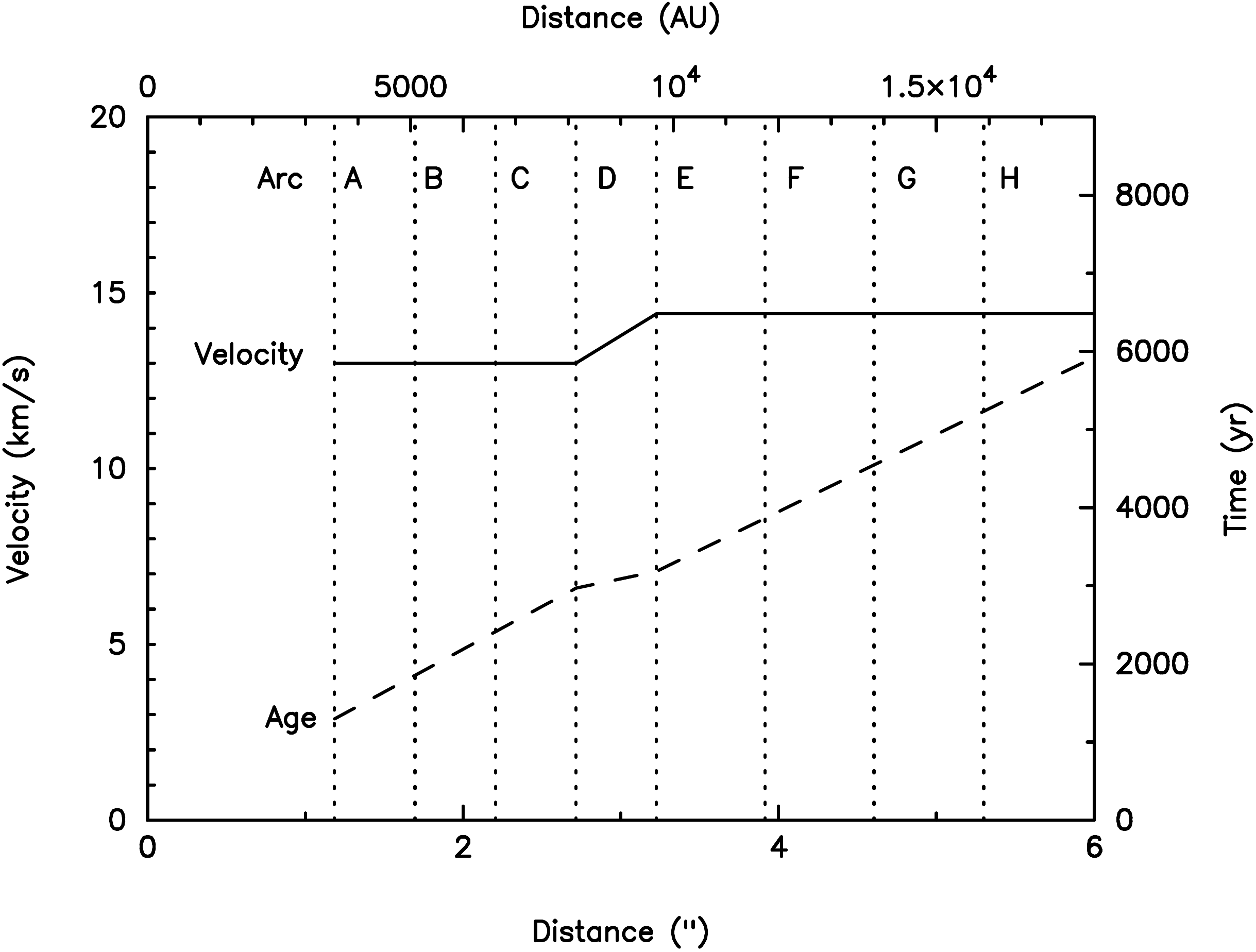}
\caption{
The expansion velocity and the dynamical age of the arcs obtained from our model (see text).
The expansion velocity decreases slightly from 14.4 to 13 km s$^{-1}$ from arc E
to arc D.   
\label{F9}}
}
\end{figure*}

The expansion velocity can be determined from the semi-major axis of the
(half) elliptical PV structure.  Unfortunately, the semi-major axis of the
absorption stripes can not be clearly identified from the observed PV
diagram.  Since the separation between two consecutive arcs is roughly the
same for the inner arcs but increases from arc D to arc E (see Figure
\ref{F8}b), we assume the expansion velocity to be the same for the inner
arcs and to increase from arc D to arc E.  From modeling the absorption
stripes with this assumption, the expansion velocity of inner arcs A, B, C,
and D are found to be $\sim$ 13 km s$^{-1}$, and the expansion velocity of
the outer arcs E, F, G, and etc are found to be $\sim$ 14.4 km s$^{-1}$,
with a small velocity jump from arc D to arc E, as shown in Figure \ref{F9}.
As can be seen, the resulting model absorption stripes (as indicated by the
dashed curves) can match the observed absorption stripes reasonably well
(Figure \ref{F8}a).  Further observations at higher sensitivity are 
needed to derive a precise value of the velocity jump from the PV diagrams.

The dynamical ages of arcs A, B, C, D, E, F, G, and H are estimated to be
$\sim$ 1300, 1850, 2410, 2970, 3180, 3860, 4550, and 5230 yr, respectively
(see Figure \ref{F9}), using the distance and expansion velocity of the arcs
in our model.  Thus, the time durations of the inner and outer arcs are
$\sim$ 550 and 680 yr, respectively.  Hence, the AGB wind velocity appeared
to decrease slightly from 14.4 to 13 km s$^{-1}$ at $\sim$ 3200 yr ago.

\section{Model for I17150} \label{MD}

\begin{figure*}
\centering{
\vspace{-4cm}
\includegraphics[angle=0,width=8cm]{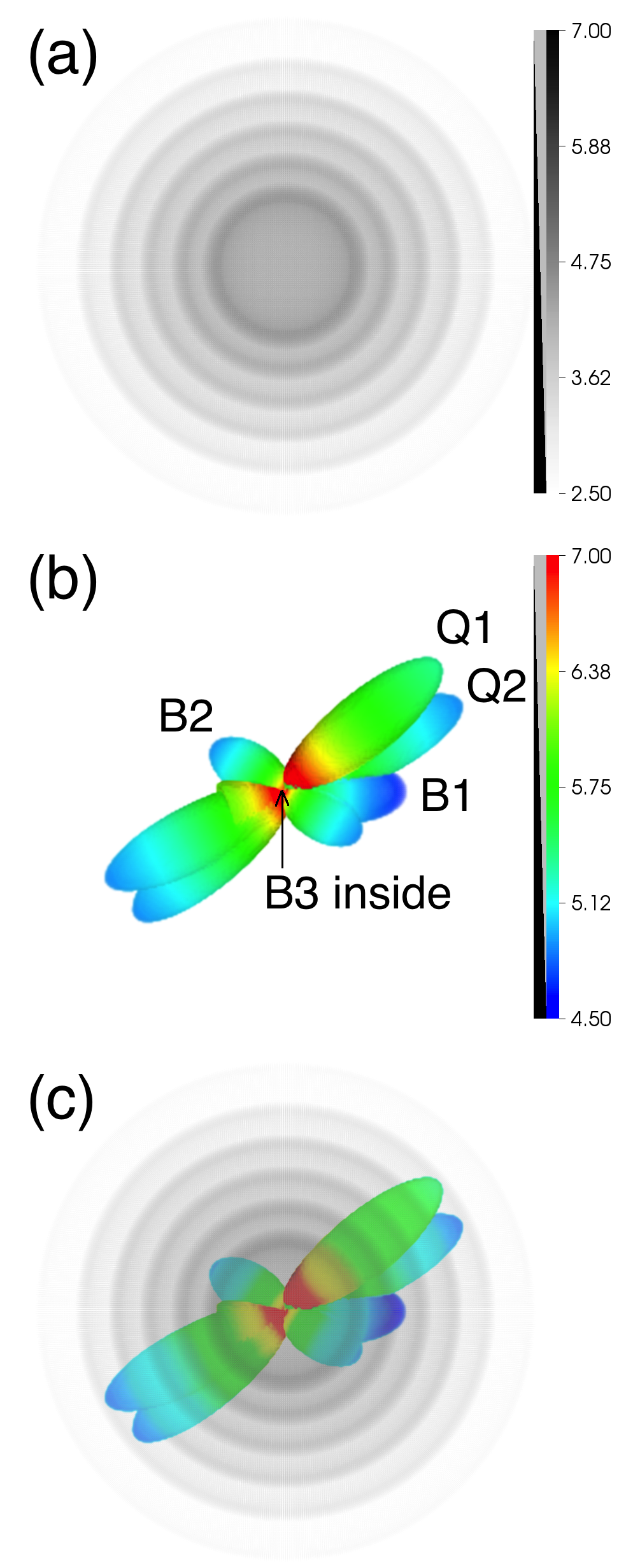}
\caption{
Our AGB wind and outflow model for I17150.
(a) shows the AGB wind.  The semi-transparent grey color 
represents the number density of H$_2$ in log scale.  (b) shows
the outflows.  The colors represent the number density of H$_2$ in 
log scale.  (c) shows the combination of the two.
\label{F10}}
}
\end{figure*}

Here we derive the physical properties (including temperature, density,
velocity, and the mass-loss rate) of the AGB wind and the outflows by
modeling their observed intensity maps, line profile, and PV structures in
CO $J=3$--2.  The aim is to determine the mass-loss process in the end of
the AGB phase and the shaping mechanism of the outflows.  Our model 
consists of a spherical AGB wind, a quadrupolar outflow, and three pairs 
of bipolar outflows, as shown in Figure \ref{F10}.  The physical 
properties of the AGB wind and outflows in our model are described as follows.

\subsection{AGB Wind}

Based on the optical arcs observed in the $HST$ image, the AGB wind is
assumed to be spherical with a periodical density enhancement.  Thus, the
density of the AGB wind can be written in spherical coordinates as

\begin{equation}
\rho_\textrm{\scriptsize a}(r) = \frac{\dot{M}_{\textrm{\scriptsize a}}(r)}{4 \pi r^2\, v_\textrm{\scriptsize a}(r)}, 
\end{equation}

\noindent where $r$ is the radial distance to the center, and
$\dot{M}_{\textrm{\scriptsize a}}(r)$ is a $r$-dependent mass-loss rate to
account for the periodical density enhancement.  The variation amplitude in
mass-loss rate is unknown.  Here we use the following simple form for the
first attempt,

\begin{equation}
\dot{M}_\textrm{\scriptsize a}(r) = \left \{ \begin{array}{ll}
    2 \, \mathrm{cos}^{2}( \pi \left(r-r_\textrm{\tiny A}\right)/\Delta r_1) \dot{M}_\textrm{\scriptsize {a,0}}, 
    \quad \mbox{if $r \leq r_\textrm{\tiny E}$} \\
    2 \, \mathrm{cos}^{2}( \pi \left(r-r_\textrm{\tiny E}\right)/\Delta r_2) \dot{M}_\textrm{\scriptsize {a,0}}, 
    \quad \mbox{if $r > r_\textrm{\tiny E}$} \\
    \end{array} ,\right.
\end{equation}
where the averaged AGB mass-loss rate 
$\dot{M}_{\textrm{\scriptsize {a,0}}}=5.3\times10^{-4}$ M$_\sun$ yr$^{-1}$,
as in the model of \citet{M02}.  
As estimated from the $HST$ image,
$r_\textrm{\tiny A}$ and $r_\textrm{\tiny E}=$ 
3555 and 9675 au, respectively.  The arc separations $\Delta r_1$ 
for the inner arcs and $\Delta r_2$ for the outer arcs are $\sim$ 1530 au 
(\arcsa{0}{51}) and 2070 au (\arcsa{0}{69}), respectively.  
The density crests are located at the radii of the arcs, where 
$r=r_\textrm{\tiny A}$, $r_\textrm{\tiny A}+\Delta r_1$, ...,  
$r_\textrm{\tiny E}$, $r_\textrm{\tiny E}+\Delta r_2$, ..., etc.

The expansion velocity of the AGB wind has been estimated earlier by
modeling the absorption stripes in the PV diagram.  It can be written with
the following form

\begin{equation}
v_\textrm{\scriptsize a}(r) = \left \{ \begin{array}{lll}
    13 \quad \rm{km} \, \rm{s}^{-1},  \quad \mbox{if $r \leq r_\textrm{\tiny D}$} \\
    13+1.4\frac{r-r_{\rm\tiny D}}{r_{\rm\tiny E}-r_{\rm\tiny D}} \quad \rm{km} \, \rm{s}^{-1},  
     \quad \mbox{if $r_\textrm{\tiny D} < r < r_\textrm{\tiny E}$} \\ 
    14.4 \quad \rm{km} \, \rm{s}^{-1}, \quad \mbox{if $r \geq r_\textrm{\tiny E}$} \\
    \end{array} ,\right. 
\end{equation}

For the AGB wind temperature, we assume a power-law distribution with

\begin{equation}
T_\textrm{\scriptsize a}(r)=T_\textrm{\scriptsize {a,0}}\left(\frac{r}{r_\textrm{\scriptsize A}}\right)^{\gamma_\textrm{\scriptsize a}},
\end{equation}

\noindent
where $T_\textrm{\scriptsize {a,0}}$ is the temperature at the radius of arc A, 
and $\gamma_\textrm{\scriptsize a}$ is the temperature power-law index for the 
AGB wind.

\subsection{Outflows}

\begin{figure*}
\centering{
\includegraphics[angle=-90,scale=0.7]{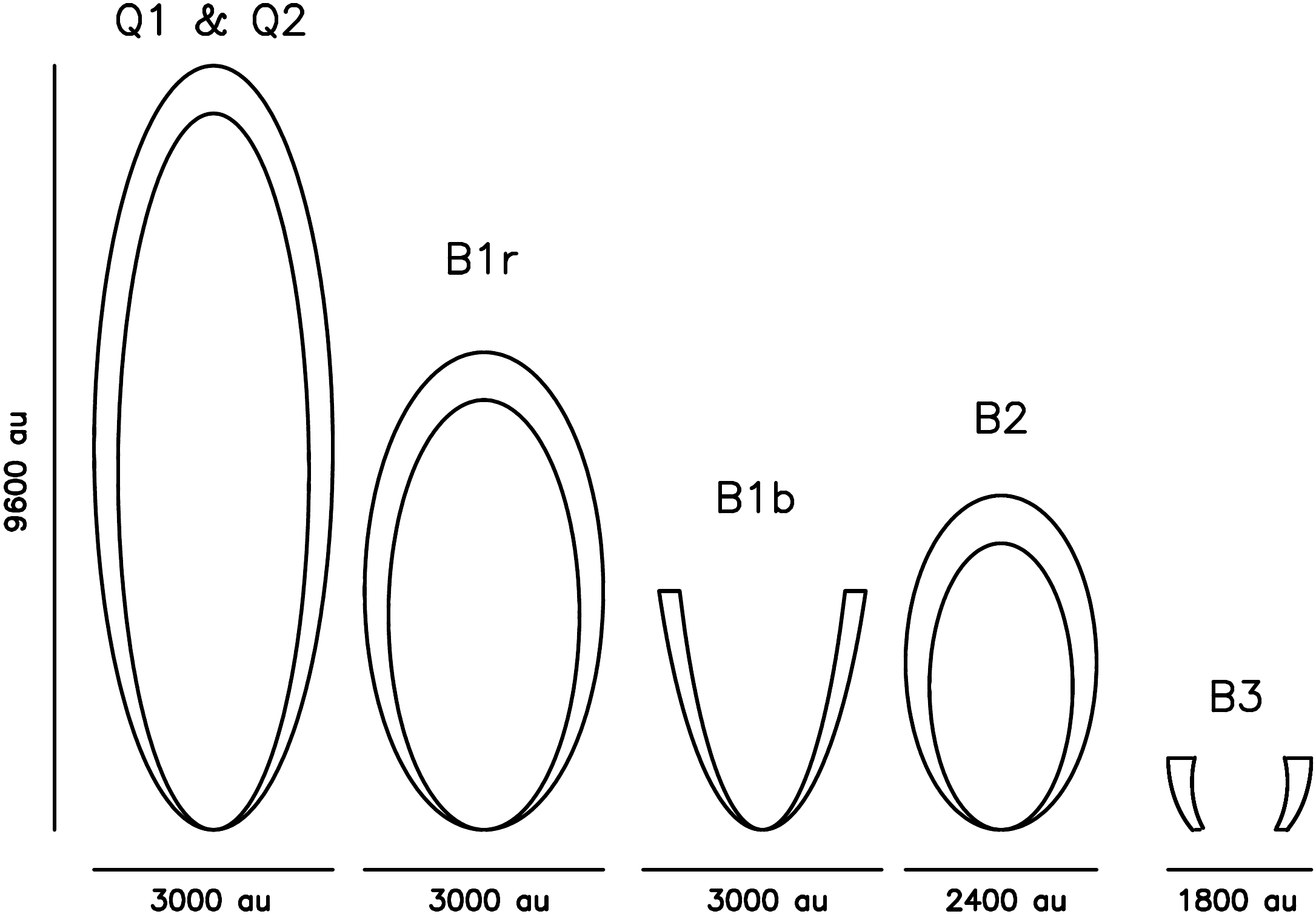}
\caption{
The outflow lobes in our model with their lengths and widths indicated.  
Each lobe is bounded by two ellipsoids/half-ellipsoids (see text).
\label{F11}}
}
\end{figure*}

For simplicity, the outflow lobes can be assumed to be ellipsoidal for Q1,
Q2, B1r and B2, and U-shaped (with a bottom half of an ellipsoid) for B1b and B3,
as shown in Figure \ref{F11}.  Note that lobe B1b is assumed to be a bottom
half of an ellipsoid because it shows a wide opening angle at the top 
and has a length only half that of lobe B1r (Figure \ref{F2}b).  Table
\ref{T2} lists the lengths $l_f$ and widths $D_f$ ($f$ is either Q1, Q2, B1,
B2, or B3) of the outflow lobes measured from the intensity-weighted
velocity map in Figure \ref{F2}b.

The thickness of the outflow shells (or cavity walls) can not be accurately
determined from the observations, because it could be smaller than the beam
size of \arcsa{0}{09} $\times$ \arcsa{0}{06}.  Following the simulation
results in, e.g., \citet{LS03}, the outflow shells here are assumed to be 
thicker at higher latitude and thinner at lower latitude. 
To construct such shells, each shell is assumed to be bounded by an outer
ellipsoid and an inner ellipsoid, as shown in Figure \ref{F11}.  In
addition, the lengths and widths of the inner ellipsoids are assumed to be
smaller than the outer ellipsoids by 600 au.  Then with the center of the
inner ellipsoid shifted by 300 au towards the central source, the shells
have a thickness increasing from 0 au at the base to 600 au at the outflow
tips.  In addition, since the bottom of outflow B3 showed an expanding
ring-like structure in the PV diagram (Figure \ref{F7}d), outflow B3 is
assumed to be two slightly merged bottom-half ellipsoids with a non-zero
width at the base.

Assuming that the outflows mostly consist of shocked or swept-up AGB wind
\citep{LS03}, their density is expected to be proportional to $r^{-2}$, as
in the AGB wind.  Thus the density of the outflow shells can be expressed by

\begin{equation}
\rho_f(r) = 1.4 \, m_{\rm{H_2}} \, n_{f,0} \, \left(\frac{r}{l_f}\right)^{-2},
\end{equation}

\noindent where $m_{\rm H_2}$ is the mass of a molecular
hydrogen, $n_{f,0}$ is the number density of molecular hydrogen at the
tips of the outflows.
Helium is also included, with its number density $n_{\rm{He}}=0.1 n_{\rm{H}}$.

The outflow emissions are mostly produced by the shocked AGB wind, and the
strongest shocks are at the outflow tips.  Therefore,
the highest temperature is expected to be found at the outflow tips. Furthermore, since we
assume that the density varies as $r^{-2}$, the temperature needs to increase with $r$, in
order to reproduce the observed ratio (roughly unity) of the emission intensity at the
outflow tips relative to that at the outflow bases.  Thus, we assume that 

\begin{equation}
T_f(r) = \left( T_{f,0}-20 \, K \right) \left( \frac{r}{l_f} \right)^{\gamma_f} + 20 \;\;\textrm{K}
\end{equation}
\noindent where $T_{f,0}$ is the temperature at the tips of the outflows, 
and $\gamma_f$ is the temperature power-law index.  

As for the outflow velocity structures, we assume that
the velocity decreases from the tips to the bases using
the following form for outflows Q1, Q2, B1, and B2:

\begin{equation}
\bm{v}_f = v_{f,0} \, {\rm{exp}} \left[- \left( \frac{\theta}{\theta_f}\right)^{\beta_f} \right] \bm{\hat{r}},
\end{equation}

\noindent where $\theta$ is the angle of the position vector $\bm{r}$
measured from the axes of outflows, $v_{f,0}$ is the velocity at the outflow
tips (where $\theta=0\degr$), $\theta_f$ is the opening angle, and
$\beta_f$ is a power-law index for the decrease of the velocity.  
Our velocity form is similar to the input velocity of the wind in \citet{LS03}. 
In general, the outflow velocity decreases from the pole to the equator as seen in the
simulations, but may decrease slower than the actual outflow velocity in the simulations.
We also do not include the converging flow seen in the simulations of the bipolar 
outflow models \citep[e.g.,][]{LS03,B19}, because the emission
detected here are mostly shocked AGB wind instead of shock wind material. \\

Unlike other outflows, outflow B3 has a non-zero expansion velocity at the
base, as implied from the PV diagram (Figure \ref{F7}d).  In addition, the
velocity of outflow B3 can decrease differently in $z$ and $R$
directions.  Thus, we assume separate equations for the two directions:

\begin{equation}
\bm{v}_{z} = v_{\scriptscriptstyle\rm B3z,0} \, \left( \frac{90\degr - \theta}{90\degr - \theta_{\scriptscriptstyle\rm B3}}\right) \bm{\hat{z}},
\end{equation}
\begin{equation}
\bm{v}_{R} = v_{\scriptscriptstyle\rm B3R,0} + \left(v_{\scriptscriptstyle\rm B3R,1}-v_{\scriptscriptstyle\rm B3R,0}\right) \, \left( \frac{\theta - \theta_{\scriptscriptstyle\rm B3}}{90\degr - \theta_{\scriptscriptstyle\rm B3}}\right) \bm{\hat{R}},
\end{equation}

\noindent where $\sqrt{v_{\scriptscriptstyle\rm
B3z,0}^2+v_{\scriptscriptstyle\rm B3R,0}^2}=v_{\scriptscriptstyle\rm B3}$ is
the velocity at the tops of outflow B3 [where
$\theta=\theta_{\scriptscriptstyle\rm B3}=\arctan
\left(0.5 D_{\scriptscriptstyle\rm B3}/l_{\scriptscriptstyle\rm B3}\right) $], 
and $v_{\scriptscriptstyle\rm B3R,1}$ is the velocity at the base.

\subsection{Model Results}

Tables \ref{T1} and \ref{T2} list the best-fit parameters in our model.  The
CO $J=3$--2 emissions are calculated using a radiative transfer code with an
assumption of local thermal equilibrium \citep{H16}.  In addition, we assume
a CO abundance of $2 \times 10^{-4}$, similar to that of the circumstellar
envelopes of O-rich stars \citep{W05}.

\begin{figure*}
\centering{
\includegraphics[angle=-90,scale=0.7]{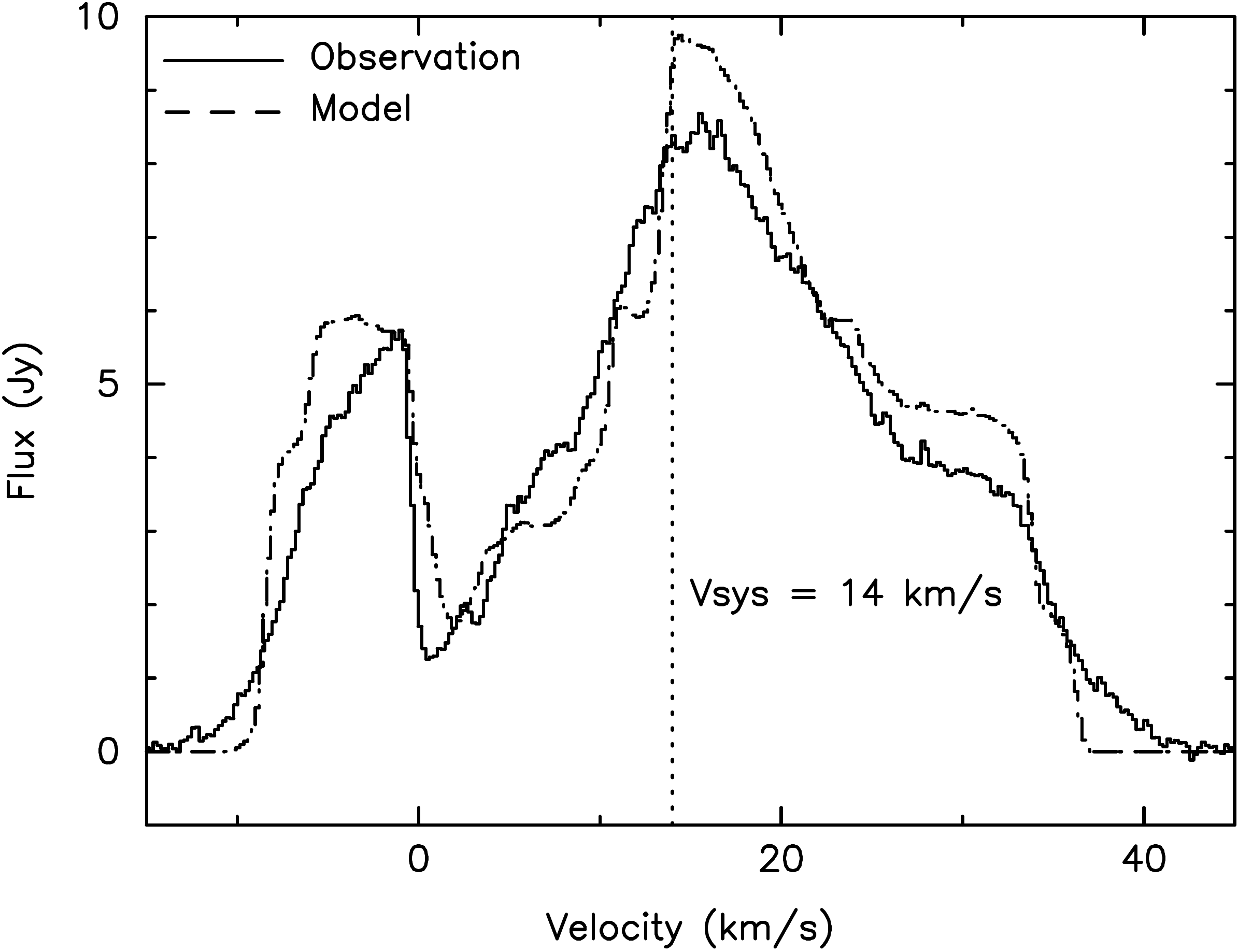}
\caption{
Comparison of the CO $J=3$--2 line profile in I17150 between our model (dashed line) and
the observation (solid line).
\label{F12}}
}
\end{figure*}

\begin{figure*}
\centering{
\includegraphics[angle=-90,scale=0.7]{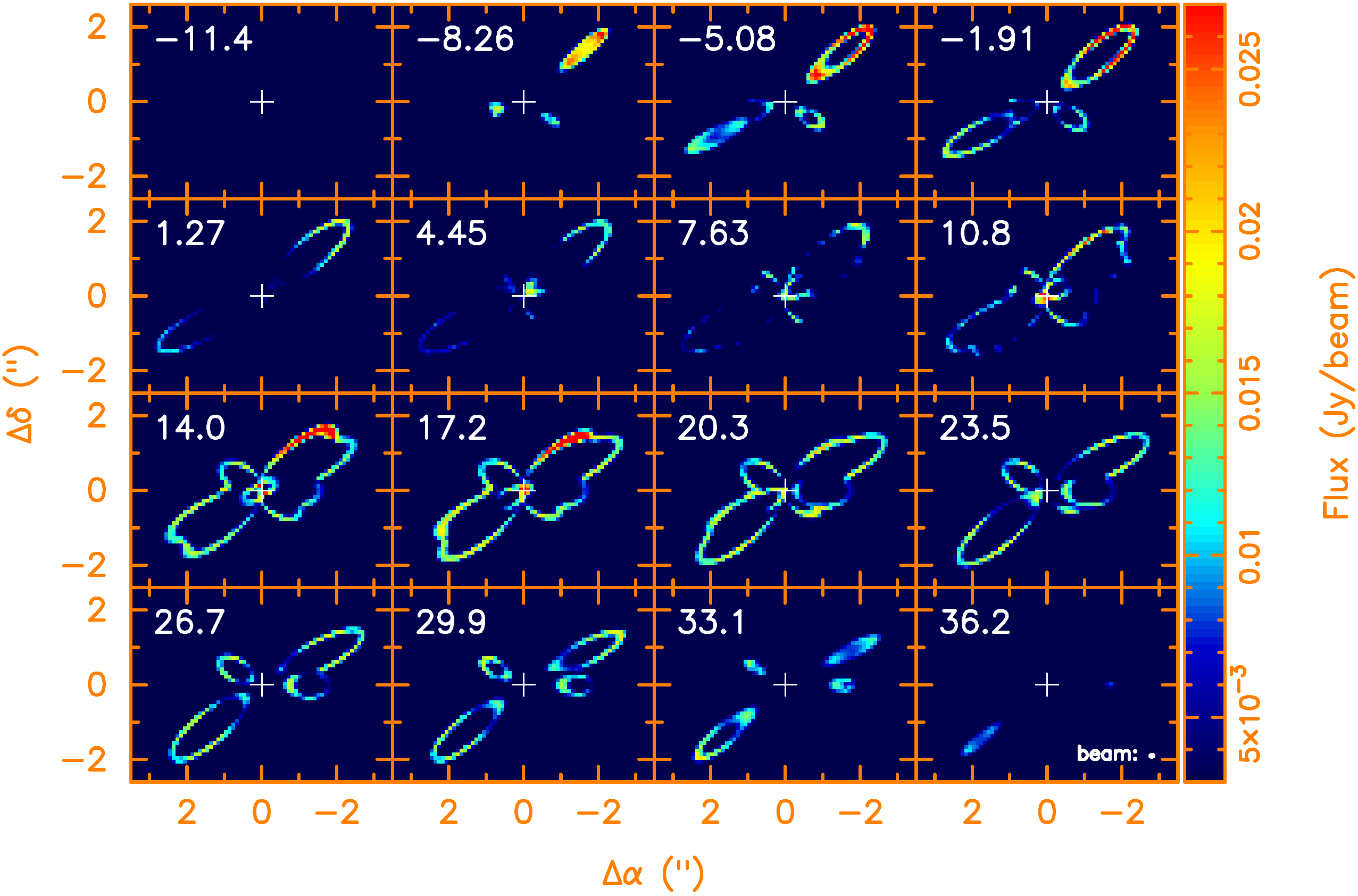}
\caption{
Simulated CO $J=3$--2 channel maps to be compared with the observations.  
The beam is the same as that in Figure \ref{F4}.  
\label{F13}}
}
\end{figure*}

\begin{figure*}
\centering{
\includegraphics[angle=-90,scale=0.7]{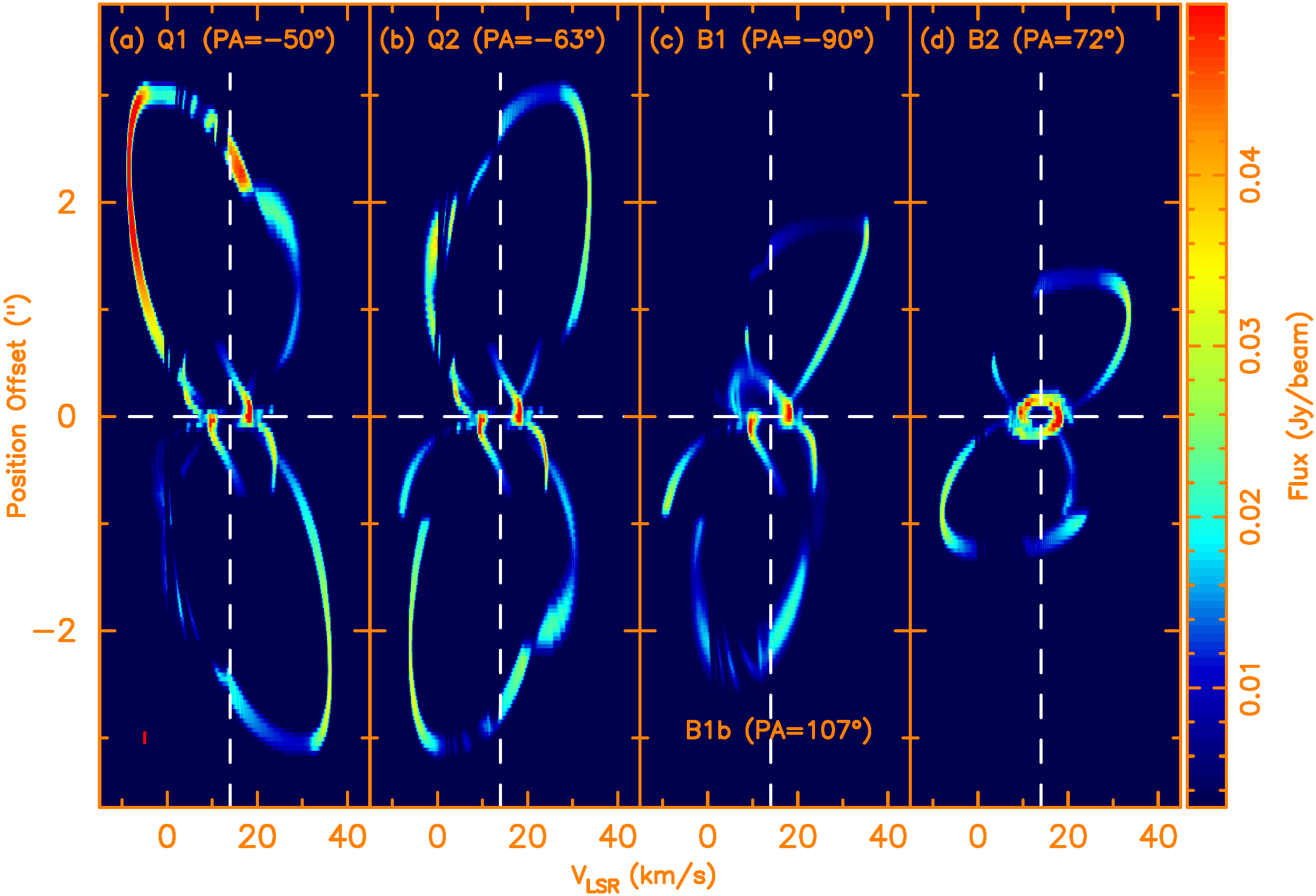}
\caption{
Simulated PV diagrams of CO $J=3$--2 emissions cut along the axes of
outflows Q1, Q2, B1 (B1r and B1b), and B2. 
\label{F14}}
}
\end{figure*}

The following criteria are used to judge the goodness of our model: (1) In
the CO $J=3$--2 line profile, the model flux intensity peaks (one on the
blueshifted side, one near the systemic velocity, and one on the redshifted
side) and the absorption dip (between $-$1 to 12 km s$^{-1}$) should be
consistent with the observed values within 15\%, as shown in Figure
\ref{F12}.  (2) In channel maps (Figure \ref{F13}), the model morphology
(including the outflow lengths, widths, position angles, etc.) and the flux
intensity should be consistent with those measured from the observed channel
maps (Figure \ref{F4}).  A minor difference can come from the high-velocity
emissions at the outflow tips (e.g., in the first and last channels of the
channel maps), because we assume outflows Q1, Q2, and B2 (except outflow B1)
are symmetric in our model, but the observed outflow velocity is slightly
asymmetric.  (3) The model PV structures along the outflow axes Q1, Q2, and
B2 should appear as tilted ellipses (Figure \ref{F14}), as in the observed
PV diagrams (Figure \ref{F7}).  Moreover, the curvatures and velocity ranges
of the model PV structures and the velocity at outflow tips should be
consistent with the observations.  (4) The outflow shells should be
partially absorbed by the front part of the AGB wind, producing broken PV
structures as in the observed PV diagrams.  The periodic density enhancement
of the AGB wind should produce the absorption stripes in the PV diagrams
(Figure \ref{F14}), with the outflow emissions fully absorbed by AGB
material in the arcs and partially absorbed by AGB material in between arcs. 
The AGB wind mass-loss rate of $5.3\times10^{-4}$ M$_\sun$ yr$^{-1}$ 
appears to be reasonable because the model spectrum can fit the observed one 
as shown in Figure \ref{F12}. 

Note that unlike the observed PV diagrams (Figure \ref{F7}), we do not 
see any faint emission inside the elliptical PV structures (Figure \ref{F14}), 
likely because we do not include any tenuous gas inside the outflow lobes 
in our model.

In our model, the velocities at the tips of outflows Q1, Q2, B1, B2, and B3
are $\sim$ 130, 130, 100, 100, and 80 km s$^{-1}$, respectively.  The
resulting dynamical ages are $\sim$ 350 yr for outflows Q1 and Q2, 280 yr for
outflow B1, 200 yr for outflow B2, and 50 yr for outflow B3.  The total mass 
of the outflows are $\sim$ 0.6 M$_\sun$.

\section{Discussions}

\subsection{New Features of I17150}

We have identified four new features in I17150.  Firstly, although this
object shows a bipolar morphology in the optical, it shows a quadrupolar
outflow in CO, with two overlapping bipolar outflows.  These two bipolar
outflows partially merge, appearing as a single bipolar outflow as seen in
the optical.  These two bipolar outflows have opposite inclinations to the
plane of the sky, as found in the velocity map.  They both have a dynamical
age of $\sim$ 350 yr.

Secondly, the quadrupolar outflow is followed by two more younger bipolar
outflows ejected along two different directions.  One is along the B1 axis
in between the symmetry axis of the quadrupolar outflow and the equatorial
plane, with a dynamical age of $\sim$ 280 yr.  Another one is almost on the
equatorial plane and has a dynamical age of $\sim$ 200 yr.

Thirdly, a bipolar outflow, B3, is seen near the central source,
appearing as a U-shaped cavity wall on each side of the source. It
may consist of shocked toroidal envelope resulting from an
interaction between the torus and an underlying wind, jet, or bullet coming
from the central source, as seen in \citet{H16}.

Fourthly, a series of absorption stripes are detected in the CO PV diagrams,
associated with the optical arcs.  These arcs trace a series of spherical
expanding shells produced by a periodically enhanced mass loss of the AGB
wind.  These shells are cooler than the outflows and thus absorb the CO
emission of the outflows, producing the observed absorption stripes in the
PV diagrams.  By modeling the absorption stripes, we find that the expansion
velocity of the AGB wind is $\sim$ 13.0 km s$^{-1}$ for inner shells and
14.4 km s$^{-1}$ for the outer shells, and decreases slightly from 14.4 to
13.0 km s$^{-1}$ from the outer to inner shells 3200-3000 yr ago.
The underlying reason for this velocity decrease is unclear.  
It may due to a decrease in escape velocity around the central source,
which in turn can be due to a decrease in stellar mass and/or the luminosity
\citep{G94}.

\subsection{When did the AGB Phase End?} \label{ARC}

In the optical image, arc A is the innermost arc detected in the flux intensity
profile along the southern lobe.  There could be more arcs within arc A, but
their intensity could be too weak to be distinguished from that of the
optical lobes in the intensity profile.  In our observations, we also could
not determine if there are more arcs within arc A, because those arcs,
even if present, could not fully absorb the strong CO outflow emission there
to produce clear absorption stripes.

In our model, arc A was ejected at $\sim$ 1300 yr ago.  If there is no more
arc ejected after arc A, then the time delay between the end of the AGB wind
and the start of the outflow would be $\sim 1300-350=950$ yr, which seems
too long comparing to the previously estimated time delay (with a median of
300 yr) between the jets and tori \citep{H07}.  However, if there was
indeed one more arc ejected after arc A, that arc would be ejected $\sim$
750 yr ago, assuming the same ejection period of 550 yr as the other inner arcs.  In
that case, the time delay would be $\sim 750-350=400$ yr, well consistent
with the previously estimated time delay.  On the other hand, if the AGB
wind continued until the ejection of the quadrupolar outflow, there would be
no time delay between the end of the AGB wind and the start of the outflow,
so the source of the AGB wind and the sources of the outflows would be
different.  For example, the AGB wind and the outflows could be ejected from
the primary star and the companion, respectively, in a binary system.  More
observations are needed to check which scenario is correct.

\subsection{Possible Shaping Mechanisms for the Quadrupolar Outflow}

The quadrupolar outflow consists of two bipolar outflows Q1 and Q2 with the
same size and dynamical age, suggesting that they are produced
simultaneously.  Many scenarios have been proposed to produce bipolar
outflows \citep{H07}: a magnetic wind from a single star or a binary; an
accretion disk around a binary companion or a primary.  If the quadrupolar
outflow was shaped by two bipolar winds, each bipolar wind should be ejected
from one star.  However, the chance of simultaneously producing two bipolar
winds from two stars is very low.  On the other hand, like the multipolar
outflows seen in other objects, e.g., in CRL 618 \citep{S04,L13a}, the
quadrupolar outflow could be shaped either by a precessing collimated fast
wind with time-dependent ejection velocity \citep{M12,V12,V13,R14} or by 
multidirectional bullets \citep{M06,B13,H16}.  If the two bipolar outflows were 
shaped by a precessing collimated fast wind, they should have different 
dynamical ages. Thus, the quadrupolar outflow seems to be shaped by 
multidirectional bullets.  

Multidirectional bullets could come from an explosive event, e.g., a
magneto-rotational explosion or a nova-like explosion
\citep{M06,B13,L13b,H16}.
In the case of the magneto-rotational explosion,
the central star spins so fast that the magnetic field above the surface of
the star is highly twisted.  Thus, the magnetic pressure force could drive
the magneto-rotational explosion, ejecting bullets (clumps of mass) in many
directions.  
In the 2nd case, a nova-like explosion is triggered by the
accretion of gas from a primary onto a white dwarf companion.  The hydrogen
gas can be pulled onto the surface of the white dwarf and then form an
envelope massive enough to ignite a hot CNO cycle in a nova explosion 
\citep{W10}.  

In our observations, the central region could indeed harbor a
binary system, with the continuum map showing a primary peak at the center
of the quadrupolar outflow and a secondary peak in the west along the
equatorial plane.  It is possible that the star at the secondary peak is
losing a part of its mass to the star at the primary peak, forming an
accretion disk and torus around the star there.

In previous study, IRAS 19475+3119 was also found to be a quadrupolar PPN
proposed to be shaped by two intrinsically collimated winds or jets
\citep{HL11}, but the launching mechanism for the winds or jets is still
unclear.  If the quadrupolar outflow lobes of IRAS 19475+3119 have the same
dynamical age, they could also be produced by multidirectional bullets.

\subsection{A Precessing Collimated Fast Wind Model for the Younger Outflows}

In addition to the quadrupolar outflow, we have also detected two younger
bipolar outflows B1 and B2 at lower latitudes, with B2 along the equatorial
plane and B1 in between the polar axis and equatorial plane.  Similar low 
latitude outflows or equatorial outflows have been observed in other PPNs, 
such as Egg Nebula \citep{S98,C00,B12}, Frosty Leo Nebula \citep{S00}, and 
CRL 618 \citep{C03,L13a}. 

According to our model, the two low-latitude outflows were ejected 280 and
200 yr ago, respectively, both younger than the quadrupolar outflow (350
yr).  Since these two outflows are younger than the quadrupolar outflow, it
is not clear if they can be produced by the same explosive bullets as for
the quadrupolar outflow.  
Unlike the quadrupolar outflow lobes produced by bullets with the same 
dynamical age, the two low-latitude outflows with different dynamical ages 
could be produced by different mechanism.  Interestingly, for these outflows, 
their dynamical ages decrease with their latitudes: Q1 = Q2 $>$ B1 $>$ B2, 
probably suggesting that the outflow axes have changed from high latitude to 
lower latitude in $\sim$ 150 yr.  
But there is insufficient data to determine the dynamical ages for other objects 
with multipolar outflows at low latitude.  

A rapid change of the outflow direction has been proposed in the precessing jet 
model for PPN CRL 618 \citep{R14} and the binary model for bullet ejections of 
the carbon star V Hydrae \citep{S16}.  In these two models, the jet and bullets 
are ejected near the periastron passage of a binary companion orbiting in a 
highly eccentric orbit \citep{SM13}.  In addition, a precessing jet with a 
time-dependent ejection velocity could produce non-overlapping lobes \citep{R14}.  
If the outflow source of I17150 is also  
a binary, the orbital period and velocity variation period can be obtained from 
the time durations between the outflows, which are $\sim$ 70--80 yr.  The 
precessing period of the disk would be two times the time duration between the 
quadrupolar outflow and the equatorial outflow (assuming that they were produced 
at opposite directions).  The precessing period is thus $\sim$ 300 yr, which is 
four times the orbital period as in the precessing jet model of \citet{R14}.  

\begin{figure*}
\centering{
\includegraphics[angle=0,scale=0.5]{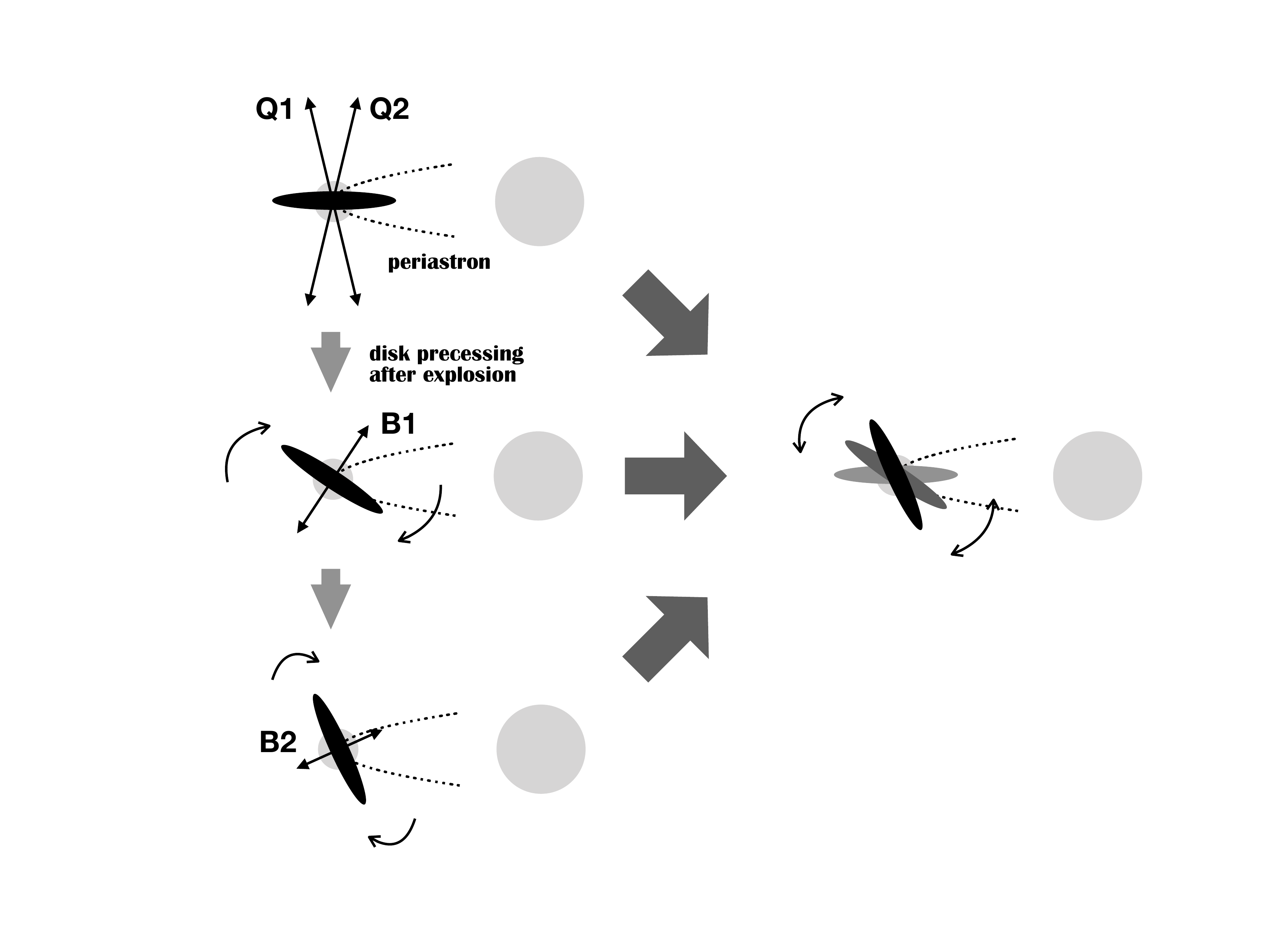}
\caption{
Schematic diagram of the ejection period.   At the first stage, 
the disk plane was on the orbital plane before the quadrupolar outflow 
was produced by an explosive event when the companion passed the 
periastron.  At the second stage, the disk started to precess due to the 
explosion and the outflow axis was changing with time.  When the companion 
passed through the periastron again, it produced outflow B1.  The third 
stage was similar to the second stage, except the larger inclination angle 
from the oribtal plane, and the outflow B2 was produced when the companion 
passed through the periastron.
\label{F15}}
}
\end{figure*}

One possibility of the outflow formation mechanism is shown in Figure \ref{F15}.  
At the beginning, the explosion, 
either magneto-rotation explosion or nova-like explosion, produced the bullets 
creating the quadrupolar outflows Q1 and Q2.  Then, the disk precessed 
by $\sim$ 35\degr\ to produce the B1 lobes, and then 
precessed more by $\sim$ 30\degr\ to produce the B2 lobes.  Producing this large 
precessing of the disk would require a binary companion with an orbital
plane significantly misaligned with the disk plane \citep{T98}.

\section{Summary}

We have mapped the AGB wind and outflows in I17150 with ALMA at $\sim$
\arcsa{0}{09} resolution.  A continuum source is detected with a primary
peak at the center of the outflows.  It is elongated along the equator,
likely tracing a dusty torus around the central source.  It also has a
secondary peak at $\sim$ 550 au to the east along the equatorial plane, and
further observation is needed to study its origin.  Two additional continuum
emissions are also detected on either side of the central continuum source,
probably resulting from interactions of CFWs with the envelope along the
equator.

Along the optical lobes, two overlapping bipolar outflows are detected in
CO, forming a quadrupolar outflow.  They have roughly the same dynamical age
of $\sim$ 350 yr.  They could be produced by bullets from an explosion such
as a magneto-rotational explosion or a nova-like explosion.  Two younger
bipolar outflows are detected with one at a lower latitude ejected 280 yr
ago and one along the equator ejected 200 yr ago.  They could be produced by
a precessing CFW with time-dependent ejection velocity.  
The PV structures of all the outflows appear as tilted
ellipses, consistent with a wind-driven shell model.  In addition, a pair of
U-shaped cavity walls are detected at the center, probably tracing shocked
toroidal envelope resulting from  an interaction with a bipolar wind.

Absorption stripes are detected in the PV diagrams of the outflow emissions
in CO, due to the absorption by cooler AGB shells in front of the outflows. 
By modeling the absorption stripes, we find that the expansion velocity to
be $\sim$ 14.4 km s$^{-1}$ for the outer shells and 13 km s$^{-1}$ for the
inner shells.  The ejections of the AGB wind could have ended after the
ejection of the innermost shell at $\sim$ 1300 yr ago.
The time delay between
the AGB wind and the outflows is $\sim$ 1000 yr, but it could be shorter
if there were additional AGB shell within the detected innermost shell.

\acknowledgements This paper makes use of the following ALMA data:
ADS$/$JAO.ALMA{\#}2016.1.01530.S.  ALMA is a partnership of ESO
(representing its member states), NSF (USA) and NINS (Japan), together with
NRC (Canada), MOST and ASIAA (Taiwan), and KASI (Republic of Korea), in
cooperation with the Republic of Chile.  The Joint ALMA Observatory is
operated by ESO, AUI$/$NRAO and NAOJ.  We acknowledge grants from the
Ministry of Science and Technology of Taiwan (MoST 104-2119-M-001-015-MY3
and 107-2119-M-001-040-MY3) and the Academia Sinica (Investigator Award
AS-IA-108-M01).

\appendix

\clearpage

\clearpage

\clearpage

\clearpage

\clearpage

\clearpage

\clearpage

\clearpage

\clearpage

\clearpage

\clearpage

\clearpage

\clearpage

\clearpage

\clearpage

\clearpage

\begin{table}

\caption{Best-fit Values of Model Parameters for the AGB Wind\label{T1}}
\begin{tabular}{cc}
\tableline
\tableline
Radius & \\
\tableline
$r_\textrm{\tiny A}$ & 3555 au \\
$r_\textrm{\tiny B}$ & 5085 au \\
$r_\textrm{\tiny C}$ & 6615 au \\
$r_\textrm{\tiny D}$ & 8145 au \\
$r_\textrm{\tiny E}$ & 9675 au \\
$r_\textrm{\tiny F}$ & 11745 au \\
$r_\textrm{\tiny G}$ & 13815 au \\
$r_\textrm{\tiny H}$ & 15885 au \\
\tableline
Temperature & \\
\tableline
$T_\textrm{\scriptsize {a,0}}$ & 10 K \\
$\gamma_\textrm{\scriptsize A}$ & $-$1 \\
\tableline\tableline
\end{tabular}
\end{table}

\clearpage

\begin{landscape}
\begin{table}
\caption{Best-fit Values of Model Parameters for the Outflows\label{T2}}
\begin{tabular}{ccccccccccc}
\tableline
\tableline
Outflow lobe & $l_f$ (au) & $D_f$ (au) & i (\degr) & P.A. (\degr) & 
$v_f$ (km s$^{-1}$) & $\theta_f$ (\degr) & $\beta_f$ & $n_{f,0}$ (cm$^{-3}$) & $T_{f,0}$ (K) & $\gamma_f$ \\
\tableline
Q1b, Q1r & 9600 & 3000 & $-$8, 8 & $-$50, 130 & 130 & 17 & 1 & $2.3\times10^5$, $7.5\times10^4$ & 150, 80 & 0.8 \\
Q2b, Q2r & 9600 & 3000 & $-$6, 6 & 117, $-$63 & 130 & 17 & 1 & $7.5\times10^4$ & 80 & 0.8 \\
B1b, B1r & 3000, 6000 & 3000 & $-$12, 12 & 107, $-$90 & 100 & 17 & 1 & $3.8\times10^4$ & 150, 100 & 0.8 \\
B2b, B2r & 4200 & 2400 & $-$8, 8 & $-$122, 58 & 100 & 26 & 1 & $7.5\times10^4$ & 80 & 0.8 \\
B3b, B3r & 900 & 1800 & $-$6, 6 & $-$60, 120 & 80$^a$ & 45 & $-$ & $3.8\times10^4$ & 133 & -0.5 \\
\tableline
\tableline

\end{tabular}
\begin{tabular}{c}
{$a$: $v_{\scriptscriptstyle\rm B3z,0}=80$ km s$^{-1}$, $v_{\scriptscriptstyle\rm B3R,0}=9$ km s$^{-1}$, and $v_{\scriptscriptstyle\rm B3R,1}=3$ km s$^{-1}$} \\
\end{tabular}
\end{table}

\end{landscape}


\begin{thebibliography}{}
\bibitem[Balick \& Frank(2002)]{BF02} Balick, B., \& Frank, A.\ 2002, ARA{\&}A, 40, 439
\bibitem[Balick et al.(2019)]{B19} Balick, B., Frank, A., \& Liu, B.\ 2019, ApJ, 877, 30
\bibitem[Balick et al.(2012)]{B12} Balick, B., Gomez, T., Vinkovi\'{c}, D., et al.\ 2012, ApJ, 745, 188
\bibitem[Balick et al.(2013)]{B13} Balick, B., Huarte-Espinosa, M., Frank, A., et al.\ 2013, ApJ, 772, 20
\bibitem[Bujarrabal et al.(2001)]{B01} Bujarrabal, V., Castro-Carrizo, A., Alcolea, J., \& S\'anchez Contreras, C.\ 2001, A{\&}A, 377, 868
\bibitem[Bujarrabal et al.(2016)]{B16} Bujarrabal, V., Castro-Carrizo, A., Alcolea, J., et al.\ 2016, A{\&}A, 593, A92
\bibitem[Bujarrabal et al.(2018)]{B18} Bujarrabal, V., Castro-Carrizo, A., Van Winckel, H., et al.\ 2018, A{\&}A, 614, A58
\bibitem[Cox et al.(2003)]{C03} Cox, P., Huggins, P. J., Maillard, J.-P. et al.\ 2003, ApJ, 586, L87
\bibitem[Cox et al.(2000)]{C00} Cox, P., Lucas, R., Huggins, P. J., et al.\ 2000, A\&A, 353, L25
\bibitem[Davis et al.(2005)]{D05} Davis, C. J., Smith, D. M., Gledhill, T. M., \& Varricatt, W. P.\ 2005, MNRAS, 360, 104
\bibitem[Dharmawardena et al.(2018)]{D18} Dharmawardena, T. E., Kemper, F., Scicluna, P., et al.\ 2018, MNRAS, 479, 536
\bibitem[de Vries et al.(2015)]{dV15} de Vries, B. L., Maaskant, K. M., Min, M. et al.\ 2015, A{\&}A, 576, A98
\bibitem[G\'{o}rny et al.(1994)]{G94} G\'{o}rny, S. K., Tylenda, R., \& Szczerba, R.\ 1994, A{\&}A, 284, 949
\bibitem[Hsu \& Lee(2011)]{HL11} Hsu, M.-C., \& Lee, C.-F.\ 2011, ApJ, 736, 30
\bibitem[Hu et al.(1993)]{H93} Hu, J. Y., Slijkhuis, S., Nguyen-Q-Rieu, \& de Jong, T.\ 1993, A{\&}A, 273, 185
\bibitem[Huang et al.(2016)]{H16} Huang, P.-S., Lee, C.-F., Moraghan, A., \& Smith, M.\ 2016, ApJ, 820, 134
\bibitem[Huggins et al.(2007)]{H07} Huggins, P. J.\ 2007, ApJ, 663, 342
\bibitem[Justtanont et al.(1996)]{J96} Justtanont, K., Skinner, C. J., Tielens, A. G. G. M., Meixner, M., \& Baas, F.\ 1996, ApJ, 456, 337
\bibitem[Kwok et al.(1998)]{K98} Kwok, S., Su, K. Y. L., \& Hrivnak, B. J.\ 1998, ApJL, 501, B117
\bibitem[Lee \& Sahai(2003)]{LS03} Lee, C.-F., \& Sahai, R.\ 2003, ApJ, 586, 319
\bibitem[Lee et al.(2013a)]{L13a} Lee, C.-F., Sahai, R., S{\'a}nchez Contreras, C. et al.\ 2013a, ApJ, 777, 37
\bibitem[Lee et al.(2001)]{L01} Lee, C.-F., Stone, J. M., Ostriker, E. C., \& Mundy, L. G.\ 2001, ApJ, 557, 429
\bibitem[Lee et al.(2013b)]{L13b} Lee, C.-F., Yang, C.-H., Sahai, R., \& S{\'a}nchez Contreras, C.\ 2013b, ApJ, 770, 153
\bibitem[Matt et al.(2006)]{M06} Matt, S., Frank, A., \& Blackman, E. G.\ 2006, ApJL, 647, L45
\bibitem[Meixner et al.(2002)]{M02} Meixner, M., Ueta, T., Bobrowsky, M., \& Speck, A.\ 2002, ApJ, 571, 936
\bibitem[Montgomery(2012)]{M12} Montgomery, M. M.\ 2012, ApJ, 745, B25
\bibitem[Riera et al.(2014)]{R14} Riera, A., Vel{\'a}zquez, P. F., Raga, A. C., Estalella, R., \& Castrill\'on, A.\ 2014, A{\&}A, 561, A145
\bibitem[Sahai(2004)]{S04} Sahai, R.\ 2004, in ASP Conf. Proc. 313, Asymmetrical Planetary Nebulae III: Winds, Structure and the Thunderbird, ed. M. Meixner, J. H. Kastner, B. Balick, \& N. Soker (San Francisco, CA: ASP), 141
\bibitem[Sahai et al.(2000)]{S00} Sahai, R., Bujarrabal, V., Castro-Carrizo, A., Zijlstra, A.\ 2000, A{\&}A, 360, L9
\bibitem[Sahai et al.(1998)]{S98} Sahai, R., Hines, D. C., Kastner, J. H., et al.\ 1998, ApJL, 492, L163
\bibitem[Sahai et al.(2007)]{S07} Sahai, R., Morris, M., S{\'a}nchez Contreras, C., \& Claussen, M.\ 2007, 134, 2200
\bibitem[Sahai et al.(2016)]{S16} Sahai, R., Scibelli, S., \& Morris, M. R.\ 2016, ApJ, 827, 92
\bibitem[Sahai \& Trauger(1998)]{ST98} Sahai, R., \& Trauger, J. T.\ 1998, AJ, 116, 1357
\bibitem[S\'{a}nchez Contreras et al.(2018)]{SC18} S\'{a}nchez Contreras, C., Alcolea, J., Bujarrabal, V., Castro-Carrizo, A., et al.\ 2018, A{\&}A, 618, A164
\bibitem[Soker \& Mcley(2013)]{SM13} Soker, N., \& Mcley, L.\ 2013, ApJL, 772, B22
\bibitem[Terquem(1998)]{T98} Terquem, C.\ 1998, ApJ, 509, 819
\bibitem[Vel\'azquez et al.(2012)]{V12} Vel\'azquez, P. F., Raga, A. C., Riera, A., et al.\ 2012, MNRAS, 419, 3529
\bibitem[Vel\'azquez et al.(2013)]{V13} Vel\'azquez, P. F., Raga, A. C., Cant\'o, J., Schneiter, E. M., \& Riera, A.\ 2013, MNRAS, 428, 1587
\bibitem[Vel\'azquez et al.(2014)]{V14} Vel\'azquez, P. F., Riera, A., Raga, A. C., \& Toledo-Roy, J. C.\ 2014, 794, 128
\bibitem[Wiescher et al.(2010)]{W10} Wiescher, M., G\"{o}rres, J., Uberseder, E., Imbriani, G., \& Pignatari, M.\ 2010, ARNPS, 60, 381
\bibitem[Woods et al.(2005)]{W05} Woods, P. M., Nyman, L.-\r{A}., Sch\"oier, F. L., et al.\ 2005, A{\&}A, 429, 977
\bibitem[Zuckerman \& Dyck(1986)]{ZD86} Zuckerman, B., \& Dyck, H. M.\ 1986, ApJ, 311, 345
\end{thebibliography}
\end{document}